\documentclass[journal]{IEEEtran}
\usepackage{amsmath,amsfonts}
\usepackage{algorithmic}
\usepackage{array}
\usepackage[caption=false,font=normalsize,labelfont=sf,textfont=sf]{subfig}
\usepackage{textcomp}
\usepackage{multirow}
\usepackage{subcaption}
\DeclareMathOperator{\Prb}{Pr}
\usepackage{stfloats}
\usepackage{url}
\usepackage{verbatim}
\usepackage{graphicx} 
\usepackage{cite}
\usepackage{diagbox}
\usepackage[dvipsnames]{xcolor}
\usepackage{fancyhdr}

\def\BibTeX{{\rm B\kern-.05em{\sc i\kern-.025em b}\kern-.08em
    T\kern-.1667em\lower.7ex\hbox{E}\kern-.125emX}}
\usepackage{balance}

\newcommand{\copyrightnotice}{
\vspace{22cm}
\hrule height 0.4pt 
    \vspace{5pt}
    \noindent\textcopyright~2024 IEEE. Personal use of this material is permitted. Permission from IEEE must be obtained for all other uses, in any current or future media, including reprinting/republishing this material for advertising or promotional purposes, creating new collective works, for resale or redistribution to servers or lists, or reuse of any copyrighted component of this work in other works.
    \vspace{5pt}
    \hrule height 0.4pt 
    
}

\begin{document}

\twocolumn[
\copyrightnotice
]

\title{Predicting change in time production - A machine learning approach to time perception}
\author{Amrapali Pednekar , Alvaro Garrido, Yara Khaluf, Pieter Simoens
\thanks{ The authors are with IDLab, Department of Information Technology, Ghent University - imec, Belgium (Email: amrapali.pednekar@ugent.be)}
\thanks{Yara Khaluf is also with the Department of Social Sciences, Wageningen University and Research, 6706KN, Wageningen, The Netherlands}
}

\maketitle

\begin{abstract}

Time perception research has advanced significantly over the years. However, some areas remain largely unexplored. This study addresses two such under-explored areas in timing research: (1) A quantitative analysis of time perception at an individual level, and (2) Time perception in an ecological setting. In this context, we trained a machine learning model to predict the direction of change in an individual’s time production. The model’s training data was collected using an ecologically valid setup. We moved closer to an ecological setting by conducting an online experiment with 995 participants performing a time production task that used naturalistic videos (no audio) as stimuli. The model achieved an accuracy of 61\%. This was 10 percentage points higher than the baseline models derived from cognitive theories of timing. The model performed equally well on new data from a second experiment, providing evidence of its generalization capabilities. The model’s output analysis revealed that it also contained information about the magnitude of change in time production. The predictions were further analysed at both population and individual level.  It was found that a participant's previous timing performance played a significant role in determining the direction of change in time production. By integrating attentional-gate theories from timing research with feature importance techniques from machine learning, we explained model predictions using cognitive theories of timing. The model and findings from this study have potential applications in systems involving human-computer interactions where understanding and predicting changes in user's time perception can enable better user experience and task performance.

\end{abstract}


\section{Introduction}

\IEEEPARstart{T}{ime} perception, or the subjective experience of time in humans, has been extensively studied from both psychological and neuroscientific standpoints \cite{wittmann2009inner}. This research has been able to offer scientific explanations to common phrases like ``time flies when we are having fun" or ``a watched pot never boils" \cite{block1980watched}. Thus, research in time perception has shed light on the underlying mechanisms responsible for familiar phenomena, such as the accelerated sense of time during enjoyable activities and the slow passage of time during anticipation. In addition to this, timing research has helped unravel the different brain regions involved in temporal processing and the factors that can influence the subjective experience of time.

The cognitive and neural theories of time perception have advanced timing research into the realm of computational modelling. Early models of time perception such as the pacemaker-accumulator model \cite{treisman1963temporal}, attentional-gate model \cite{zakay1995attentional} and  beat frequency models \cite{miall1989storage} serve as foundational frameworks for explaining timing behaviour in humans. Building on these frameworks, researchers have integrated timing components into widely-used cognitive architectures like adaptive control of thought--rational (ACT-R) \cite{taatgen2007integrated, byrne2006act, dzaack2007computational,anamalamudi2014computational}. These studies have facilitated a better understanding of the interplay between timing and other cognitive processes in the brain. Additionally, well-documented timing biases like regression to the mean and time-order errors, have been explained through a Bayesian perspective\cite{jazayeri2010temporal, wiener2014continuous,sadibolova2022temporal}. These studies have suggested that time perception can be thought of as a probabilistic inference process in the brain, similar to other sensory perceptions. Such advances in modelling time perception have enabled the integration of timing research into application-oriented fields like robotics and human-computer interactions \cite{maniadakis2014time,maniadakis2011temporal,lourencco2020teaching}. 

While research in time perception has advanced significantly over the years, several aspects remain largely unexplored. A quantitative analysis of timing studies at an individual or sub-group level, is one such under-explored area \cite{matthews2014time}. As noted by Matthews\cite{matthews2014time}, most timing research studies aggregate the data across participants, reporting findings at a population level but ignoring individual differences in timing behaviour. Thus, well-established timing effects observed at the population level may not hold for specific individuals or sub-groups. This limitation extends to the computational models of timing. These models are often evaluated for their ability to qualitatively replicate human timing behaviour at a population level and usually lack output analysis at an individual or sub-group level. 
In addition to analysing individual level data, a quantitative approach can enable the use of predictive modelling techniques like machine learning on time perception data. Thus, it can open new avenues to not only analyse individual differences in timing behaviour but also predict time perception of an individual given a set of conditions.

Another under-examined area in timing research is, time perception in an ecological (or naturalistic) setting~\cite{matthews2014time}, \cite{van2018towards}.  A majority of timing studies rely on simple stimuli like flashes, tones and images, with a small number of participants performing synthetic tasks that do not resemble real-world scenarios. As argued by several authors \cite{matthews2014time}, \cite{van2018towards}, findings and models derived from such experiments may not be applicable in real-world settings. Thus, there is a need to move towards more complex stimuli and have a larger pool of participants perform slightly more realistic tasks. Such an approach to studying time perception in naturalistic settings would in turn reinforce the need to analyse individual and sub-group differences, as discussed above. This is because, naturalistic settings introduce greater variability in environmental and individual characteristics, potentially amplifying individual deviations in timing behaviour from the population mean. Consequently, an ecological approach to time perception goes alongside accounting for quantitative differences at an individual or sub-group level \cite{van2018towards}.

Addressing these gaps in timing research is important because, in addition to understanding the timing mechanisms in humans, time perception research is gaining importance from an application standpoint in fields like robotics and human robot interactions \cite{maniadakis2011temporal}. For example, ChronoPilot \cite{botev2021chronopilot} is a system that uses extended reality technologies to engage with a user performing a task. The aim of this system is to enable well-being and better task performance by modulating the subjective time experience of a user. Another example, as described in \cite{maniadakis2017emotionally}, involves robots assisting multiple users. The overall user experience of such a system is improved by the robot prioritizing users with a faster perceived time, thus optimizing the perceived wait times of its users. Such applications highlight the importance of understanding an individual's time perception in naturalistic settings. Therefore, a quantitative and ecologically valid approach to timing research can open new avenues for both theoretical research and practical applications.

One potential obstacle for an application-oriented approach is maintaining a balance between model performance and model explainability (especially, within the framework of existing theories of time perception). Models grounded in cognitive architectures provide explanations consistent with cognitive theories of timing \cite{taatgen2007integrated, byrne2006act , dzaack2007computational}. However, these models often show poor quantitative performance, especially in naturalistic settings, because their parameters are based on heuristics rather than real-world data distributions~\cite{durstewitz2016computational,taatgen2008constraints}. Conversely, data-driven models from machine learning tend to have a superior performance because they are trained on real-world data. However, they often function as black-box systems that give only limited explanations for their predictions~\cite{kar2022interpretability}. This suggests a need for hybrid approaches, either by incorporating statistical estimations into cognitive model parameters, or by adding cognitive explanations to machine learning model predictions.   


This study takes a small step towards an application-oriented approach to time perception. We trained a machine learning model to predict the direction of change in subjective time production of an individual in the upcoming trial of an experiment. Training data for the model was collected through an online experiment designed to study the prospective time production of participants watching a video (without audio). An online setup ensured fewer controlled variables compared to an offline experiment and allowed to collect data from a large number of participants. Additionally, the videos were simulated driving scenarios from a front-seat passenger's perspective. They were designed so as to incorporate multiple time-influencing stimuli like magnitude, salience and oddball (each of which have been mostly studied independently in laboratory settings~\cite{tachmatzidou2023attention, malpica2022larger}). Hence,  by recruiting a large number of participants (995), having fewer controlled variables, using a first-person perspective~\cite{van2014s} and combining multiple time-influencing stimuli in one dynamic stimuli like a video \cite{roseboom2019activity, riemer2021age}, we move towards a more ecological setting. An ecological setting also ensures wider applications of the model. Consequently, the machine learning model was tested on unseen data from a second experiment where it performed equally well, providing evidence for its broader applicability. An analysis of the model's output revealed that it contained information about the magnitude of change in time production in addition to direction of change. Furthermore, the features used by the machine learning model were connected to the different components of the attentional-gate model \cite{zakay1995attentional}, a cognitive framework of time perception. This enabled us to provide cognitive explanations to the predictions and features used by the model. Thus, this study makes the following contributions:

\begin{enumerate}
    \item {\bf An ecologically valid dataset of time perception:} The training data for the machine learning model consists of time production data from a large number of participants (995). This data was collected using an online experiment which ensured fewer controlled variables. The videos used in the experiment were simulated driving scenarios from a first-person perspective that combined multiple time-influencing visual stimuli. They were designed to closely resemble the experience of a front-seat passenger observing scenes from a moving car.   
    
    \item {\bf A machine learning model that predicts change in time production :} We built a machine learning model that predicts the direction of change in subjective time production of a participant between consecutive trials. The model performance is quantitatively evaluated on a held-out test set and its generalization is demonstrated on unseen data from a second slightly different experiment. Furthermore, the model output is quantitatively evaluated for it's ability to carry information about magnitude of change in time production in addition to direction of change. The model predictions have also been analysed at both the population and individual level using SHAP values \cite{scott2017unified}, a technique in explainable AI. Thus, the model and its analysis contribute towards a quantitative approach to time perception at an individual level. 

    \item {\bf Explanation of model predictions using a cognitive framework of time perception:} We connect features used by the machine learning model with different components of the attentional-gate model. Thus, by combining the attentional-gate theory of time perception with feature importance techniques of machine learning, we aim to provide a cognitive explanation to the machine learning model predictions. Thus, we were able to formulate hypotheses about reasons for change in time production for a particular participant.   

\end{enumerate}

\section{Related Work} 

\noindent In this section we discuss some previous efforts at addressing the gaps mentioned in this study.

Time perception at an individual level has been studied using physiological and neural activation data. 
In \cite{orlandic2021wearable}, the authors train a machine learning model that uses physiological biomarkers to predict whether time is passing fast or slow based on timing errors of a given individual. 
While in \cite{aust2024automatic}, the authors study the relation between physiological data and time perception in a more naturalistic environment. They train a machine learning model to predict the perceived passage of time of air traffic controllers under various workload conditions using physiological data. 
Individual differences in timing have also been studied using functional magnetic resonance imaging (fMRI). In \cite{wiener2014individual}, the authors study fMRI's of participants performing a colour and temporal discrimination task. They conclude that different regions in the brain are involved in timing, and individual differences may arise due to different levels of activities in these regions. In \cite{grahn2009neural}, the authors study the relation between difference in beat perception and difference in time perception among individuals. Using fMRI, they show that activities in auditory and motor areas in the brain is correlated with individual differences in beat perception. They further studied the relationship between the differences in activities due to varying beat perception and individual difference in time perception. 

Attempts have been made to move to a more ecological setting by using complex and dynamic stimuli in a laboratory setting. In \cite{van2014s}, the authors studied interval timing of drivers using videos generated by a driving simulator. The videos were from the drivers or co-drivers perspective with the car moving at different speeds. By using dynamic stimuli like a video with a first person perspective, the authors attempted to move closer to a real-world setting. The results confirmed existing qualitative findings from simpler settings. Namely, videos depicting faster speeds or faster moving cars are perceived to be longer compared to videos depicting slower speeds or slower moving cars. In \cite{riemer2021age}, the authors used images and videos to study the impact of naturalistic stimuli on age related differences in time perception. They conclude that due to larger information processing required for naturalistic stimuli the timing performance of younger adults is found to be better than older adults. In~\cite{roseboom2019activity}, the authors used naturalistic video recordings that combined multiple simpler visual stimuli. They studied the effect of salience on time perception and qualitatively replicated these effects using activations from different layers of a convolutional neural network. 

Apart from using dynamic stimuli in a laboratory, efforts have been made to take the experiment setup outside laboratories into the real-world where participants perform more natural timing tasks. In \cite{van2021attention}, the authors assessed the role of attention in time perception of players of an online multiplayer real-time strategy game. They found that distractions during the interval have lesser influence on time perception as compared to distractions just before the required response. They present qualitative results showing that, rather than an attention gate slowing down the cognitive counter, the variance in timing due to distraction of attention away from time is because the distraction prevents participants from checking their cognitive counter. In \cite{tobin2010ecological}, the authors study longer timing interval (from 12 minutes to 58 minutes) of video game players in a gaming centre. In this study, the authors compare prospective and retrospective timing of players. The prospective estimates were found to be longer than the retrospective ones and thus the results confirm the classical distinction found between these paradigms in a naturalistic setting. In \cite{balta2024effects}, the authors study the effect of cognitive load on time perception of air-traffic controller who were actively engaging in real flight control sessions. The results showed that higher cognitive load leads to overproduction, again confirming earlier insights of laboratory studies about the effect of cognitive load on time perception. These studies highlight the need for ecological validity in an experimental setup, to both confirm existing laboratory findings and uncover additional timing behaviours that may arise when moving from the laboratory to the real-world.

Computational modelling in time perception can be broadly divided into two types, models based on cognitive architecture and models based on Bayesian framework. Most efforts with cognitive architectures are focused around integrating the pacemaker based internal clock into the ACT-R architecture.  For example, in \cite{taatgen2007integrated}, the clock rate of the internal clock was decayed over time to ensure scalar property of time estimation \cite{gibbon1977scalar}. While in \cite{byrne2006act}, clock rate was based on attentional-gate theory and thus had a fixed mean rate. On the other hand, \cite{dzaack2007computational} added a timing module into ACT-R for retrospective timing instead of prospective timing. These models however have only been tested for their ability to qualitatively replicate tendencies and biases found in human timing data. They do not involve individual level analysis or predictions. Most efforts with Bayesian modelling are focused on explaining the popular timing biases like regression to the mean and time-order errors using a Bayesian framework \cite{sadibolova2022temporal}. Parameters of the Bayesian observer model are either based on heuristics or a combination of heuristics and human experiment data \cite{wiener2014continuous}, \cite{jazayeri2010temporal}. These studies again focus on population level timing biases and do not discuss model predictions at in individual level.  To the best of our knowledge, no studies have focused on predicting change in time perception of an individual.

Attempts to draw parallels between computational models of timing and existing theories of time perception are very few. In \cite{shi2013bayesian}, the authors discuss how the different components of a pacemaker accumulator model can be linked to the different components of the Bayesian framework. Specifically, they relate the reference memory of the pacemaker model to the prior in a Bayesian model, the clock stage to the likelihood, the working memory to the posterior and the comparator stage of the pacemaker model to the loss minimization stage in a Bayesian model. Such a comparison helps to draw parallels between a model trained on data and the existing theories of time perception. In this study, we aim to do something similar by aligning the features used by a machine learning model trained on ecologically valid timing data with the different components of the attentional-gate model of time perception \cite{zakay1995attentional}.     

\section{Methods}
In this section, we first define the methods used to move closer to an ecological setting by describing the naturalistic video stimuli and the main experiment used to collect training data for the machine learning model. Next, we outline the experiment data points used as features for the machine learning model and provide a description of the model itself. The attentional-gate model is then briefly described, with a focus on aligning the machine learning model features with different components of the attentional-gate model. This is followed by an overview of the baseline models used in the study and a description of the second experiment used to test the machine learning model’s generalization capabilities.

\subsection{Naturalistic video stimuli}

\begin{figure*}[!t]
    \centering
        \includegraphics[width=0.3\linewidth]{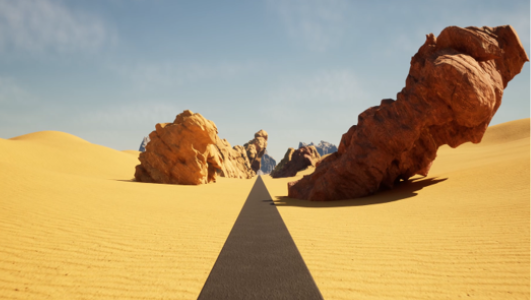}
    \includegraphics[width=0.3\linewidth]{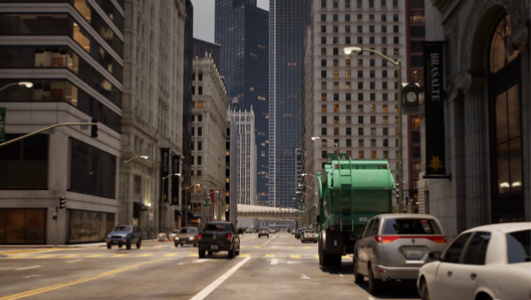}
    \includegraphics[width=0.3\linewidth]{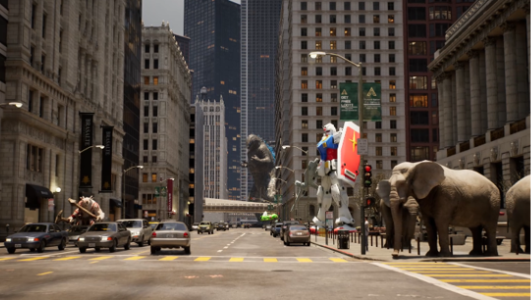}
    \caption{Screenshots of the videos used as naturalistic stimuli in the main experiment. The leftmost image corresponds to low engagement level, the middle image to medium engagement level and the rightmost image to high engagement level.}
    \label{fig:exp_stimuli}
\end{figure*} 

\noindent The video stimuli used in the main experiment consists of different simulated driving scenes from a front-seat passenger's perspective. These videos are considered naturalistic for two reasons. First, they represent a first-person perspective \cite{van2014s}, which can be considered a close resemblance to the experience of a front-passenger observing scenes from a moving car. Second, they simultaneously incorporate multiple time-influencing visual stimuli like magnitude, salience and oddball in a less controlled manner \cite{roseboom2019activity}, \cite{riemer2021age}.

``Visual engagement'' is used as an umbrella term encompassing one or more time-influencing visual stimuli. It defines the degree to which a given video captures a viewer's attention. For simplicity, videos were grouped into three levels of visual engagement (Figure \ref{fig:exp_stimuli} shows screenshots of videos under each engagement level):

\begin{itemize}
    \item \textbf{Low Engagement level}: A desert scene with sand and canyons along a straight, empty road with no other activity or objects in sight. These scenes have low salience and high predictability. Hence, they are assumed to have a low engagement level.  
    \item \textbf{Medium Engagement level}: A city scene with buildings along the two sides of a straight road, other cars moving on the road and pedestrians walking on the sidewalk. These scenes have moderate salience and predictability. Hence, they are assumed to have a medium engagement level.  
    \item \textbf{High Engagement level}: A city scene similar to the one described in medium engagement level, but with additional unusual elements. For example, a Godzilla and other animated characters from online games appear every 10 or 15 seconds along the road. These scenes have the highest salience, and induce unpredictability. Viewers may become curious about what happens next. Hence, they are assumed to have a high engagement level.  
\end{itemize}

\noindent  The videos were generated by a professional content creator using Unreal engine\cite{enwiki:1240657533}, giving us flexibility in adding or removing objects from different locations within each scene. There were a total of three videos under each type of engagement level. Links to the videos are included in the supplementary material.

\subsection{Main experiment design}

\begin{figure*}[!t]
    \centering
    \includegraphics[width=0.8\linewidth]{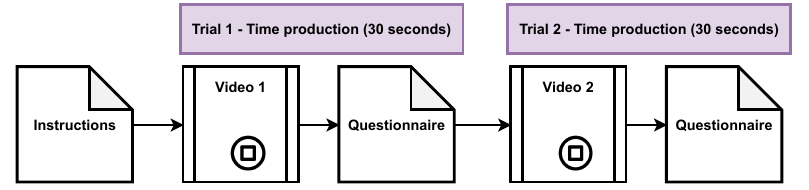}
    \caption{Main experiment design: One trial included a video followed by a questionnaire. The  two trials were time production tasks, where participants had to stop the video after they thought 30 seconds had passed. In the questionnaire following each video, participants were asked about the video content (attention checks), self-evaluation of their timing performance and their visual and timing perception }
    \label{fig:exp_setup}
\end{figure*}

\noindent The main experiment was set up to collect training data for the machine learning model. The aim of this experiment was to capture the change in prospective time perception of a participant experiencing a pure visual input (no audio or haptic).

The experiment setup is shown in Figure \ref{fig:exp_setup}. A video followed by a questionnaire, is defined as one trial.  In a trial, participants were asked to watch the video attentively and stop it when they thought 30 seconds had passed. To ensure that participants did not count, they were instructed to pay attention to the video contents in order to complete a quiz after the video. The questionnaire following each video, contained questions about the video content (attention checks), a self-evaluation of timing performance, and questions related to visual perception and passage of time. Thus, the two trials involved a prospective time production task and a non-temporal task, both of equal importance. As the target interval (30 seconds) was the same in the two trials, a change in time production in the second trial as compared to the first, was assumed to be related to change in prospective time perception. 

The video series, or the two videos shown to each participant could belong to either the same or different engagement levels. Thus, there were a total of nine permutations of the three engagement levels for the two trials. Out of these, one was selected randomly for each participant. It is important to note that each engagement level comprised of three videos. Hence, we made sure even if a participant watched the same engagement level consecutively, she would be watching a different video each time. 

This experiment was conducted on Prolific \cite{prolificProlificQuickly, douglas2023data}, an online experiment platform. The online setting enabled us to move closer to an ecological setting due to fewer controlled variables and a large number of participants. A total of 995 participants were recruited. Out of these, nine participants were excluded either because they failed the attention checks or reported technical problems during the experiment. There was no explicit criteria for gender, age or ethnicity of the participants. However, they were expected to have a good knowledge of English. The participants were paid at a rate of 8 pounds per hour (the median duration of the experiment was approximately 20 minutes). The research was approved by the Ethics committee of the Faculty of Psychology and Educational Sciences of Ghent University.

\subsection{Feature selection}

\noindent Features for the machine learning model were extracted from different data points of the experiment, such as, timing performance and questionnaire responses. Using feature elimination techniques like permutation feature importance (see supplementary information for details), the five most predictive features were selected. These features have been assigned to broader descriptive categories to enhance readability and facilitate generalization to other settings. Table~\ref{table:FeatureDescription} shows the feature categories and values for the selected  features.

\begin{table}
\begin{center}

{\renewcommand{\arraystretch}{1.3}%
{\begin{tabular}{ m{7.5em} c  m{7.5em}   } 
\hline
\textbf{Feature category} & \textbf{Feature name} & \textbf{Feature values} \\
\hline

Prior timing performance & \textit{T1RelError} & M=15, SD=44, Min=-70, Max=440  \\

 \hline

Self-evaluation of timing performance & \textit{T1LowerThan30}  & Higher than 30 = 0, Lower than 30 = 1  \\
  \hline
Participant sensitivity & \textit{HighVisualSensitivity}  & High sensitivity = 1,  others=0  \\
\hline
\multirow{2}{*}{}Environemental characteristics  & \textit{ChangeInEngagementLevel} & Lower = 0, Same = 1, Higher =2 \\
\cline{2-3}
& \textit{V2EngagementLevel} & Low=0, Medium=1, High=2  \\
\hline

\end{tabular}}
}
\captionof{table}{Feature selection: The table lists the categories, names and values of the five most predictive features selected.}
\label{table:FeatureDescription}
\end{center}
\end{table}

\begin{itemize}
    \item \textbf{Prior timing performance}: The feature \textit{T1RelError}, captures information about a participant's relative time production error. It is defined (similar to the relative time estimation error described in \cite{orlandic2021wearable}),  using the time produced in trial one, \textit{T1ProducedTime}, and the target interval, \textit{TargetInterval} (30 seconds) as follows:
    
{\small\begin{equation}
\text{\textit{T1RelError}} = \frac{\text{T1ProducedTime} - \text{\textit{TargetInterval}}  }{\text{\textit{TargetInterval}}} * 100
\label{eq:t1_production_rel_error}
\end{equation}
}

    \item \textbf{Self-evaluation of timing performance:} In the questionnaire following each video, participants were asked to categorize their produced time as either higher or lower than 30 seconds. The binary feature, \textit{T1LowerThan30}, represents a participant's corresponding response from the trial one questionnaire.

    \item \textbf{Participant sensitivity}: The feature, \textit{HighVisualSensitivity}, captures information about extreme discrepancies in reported subjective engagement level for low engagement videos. A participant is marked as sensitive, if she reports a low engagement video to be highly engaging. It is important to note that only participants who watched a low engagement video in trial one were assessed for sensitivity. All others, who watched either a medium or high engagement video in trial one, were assumed to not be sensitive by default since we could not capture an extreme discrepancy between perceived and objective engagement level for such participants. A corresponding feature that captured extreme discrepancies for high engagement videos was not included in the most predictive features. (See feature selection section in supplementary information for more information). 

    \item \textbf{Environmental characteristics}: The feature \textit{V2EngagementLevel}, represents the objective engagement level of the video presented in trial two. And, the feature \textit{ChangeInEngagementLevel}, represents the change in engagement level between the video shown in trial two and the video shown in trial one.  Both features are fully under control of the experimenter and are independent of the participant's subjective sensitivity or timing characteristics. 

\end{itemize}

\begin{figure}[!t]
    \centering
        \includegraphics[width=\linewidth]{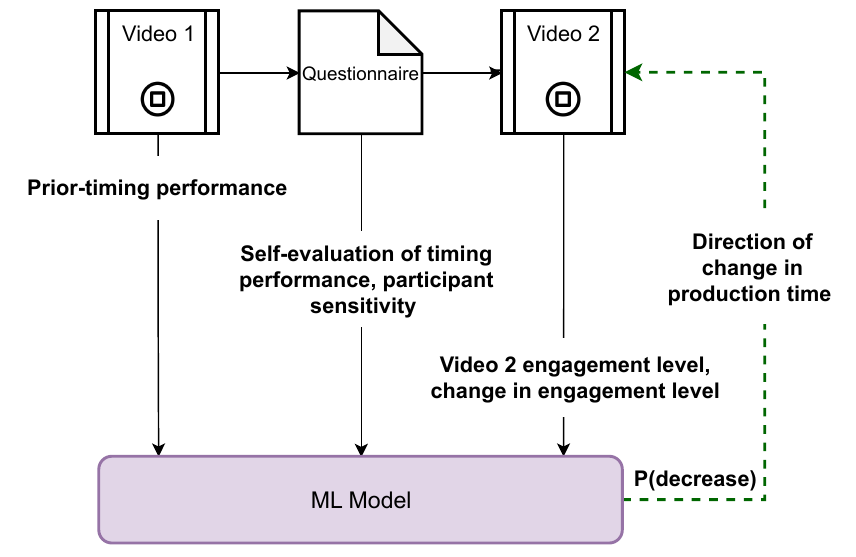}
    \caption{Prediction model: The machine learning model uses the selected features from the previous trial and the engagement level of the upcoming trial's video as inputs. Its output is a probability of decrease, which reflects the direction of change in production time. Additionally, these probabilities were found to contain information about the magnitude of change in production time (as detailed in the Results section). Using this information, we can approximately infer how long a participant might wait before stopping the video}
    
    \label{fig:prediction_problem}
\end{figure}

\subsection{Prediction model}

\noindent The prediction model is a machine learning model that uses the selected features to predict direction of change in a participant's production time. Specifically, given a set of feature values generated by a participant in the first trial and the engagement level of the upcoming trial's video, the model predicted the probability of decrease in time production in the second trial as compared to the first (see Figure \ref{fig:prediction_problem}). Thus, the model output (probability of decrease) carried information about the direction of change in time production. When the probability was more than 0.5, the direction of change was classified as decrease in time production. When it was less than 0.5, the direction of change was classified as an increase in time production. Furthermore, the model output was also found to contain some information about magnitude of change in production time. The Results section contains a detailed evaluation of the model for it's ability to convey information about both the direction and magnitude of change in time production. 

Analysis of the experiment data revealed that 36 \% of participants decrease time production in second trial as compared to the first. In order to handle this class imbalance, data was sampled using random undersampling to have equal number of increase and decrease cases. Due to this, the final training dataset had 706 participants. 

A number of classical machine learning algorithms covering different model families like tree based, probabilistic and neural networks, were tested for the prediction problem. All models had similar performances on a held-out test set. The logistic regression model was selected for its simplicity and implicit probability calibration. The supplementary information section contains further information about performance of each model and the model selection process. All algorithms were implemented using the scikit-learn library in Python. 

The following equation represents the logistic regression function used to calculate the probability of decrease, 
{\small\begin{equation}
  \Prb(\text{Y} = Decrease \mid \text{X}) = \frac{\text{1} }{1 + \text{exp}(-X)} 
  \label{eq:general_lr_eq}
  \end{equation}
}
Where X for the trained model was, 
\begin{align*}
X = 0.016
    &+ 0.662 \ *  \text{(\textit{T1RelError})} \\
    &- 0.191 \ * \text{(\textit{T1LowerThan30})} \\
    &- 0.241 \ * \text{(\textit{HighVisualSensitivity})} \\
    &- 0.187 \ * \text{(\textit{\textit{V2EngagementLevel}})}  \\
    &+ 0.177 \ * \text{(\textit{ChangeInEngagementLevel})} \\
    \label{eq:X_trained} \tag{3}
\end{align*}

Thus, each participant was represented as a set of five features and the model predicted the probability of decrease in time production. Since all feature values were scaled using scikit-learn's Standard scalar function, the coefficients in Equation \eqref{eq:X_trained} represents feature importance.  A detailed analysis about the effect of different features on the predicted probability is conducted in the Results section.

\subsection{Attentional-gate model}

\begin{figure}[!t]
    \centering
    \subfloat[][Working of attentional-gate model (simplified)]{{\includegraphics[width=\linewidth]{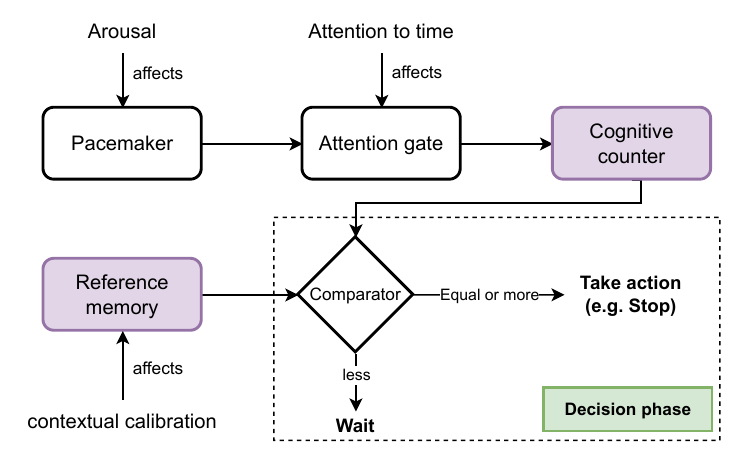}}}
    \\
    \subfloat[][Hypothesized reasons for change in time production]       {{\includegraphics[width=\linewidth]{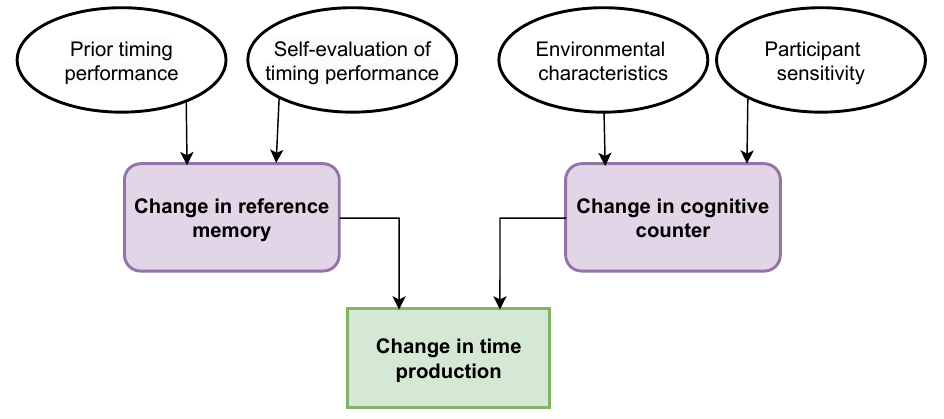}}}

    \caption{(a)  
    The cognitive counter accumulates all or fewer ticks generated by the pacemaker depending on the attention-gate width. Arousal caused by a stimuli or the environment, affects the pacemaker by increasing clock speed. Similarly, attention to time determines the width of the attention-gate component and, contextual bias like regression to the mean influences reference memory. The decision phase involves, a comparator that waits until the accumulated ticks in cognitive counter are more than or equal to the reference memory ticks in order to take an action associated with marking the end of the target interval. (b) The white ovals depict the broad feature categories used by the machine learning  model. These features can indicate a change in either reference memory or cognitive counter. We assume that self-evaluation of timing performance and prior timing performance capture information related to reference memory changes (contextual calibrations like regression to the mean and prior timing experience). While, environmental characteristics and  participant sensitivity capture information about changes in cognitive counter (arousal or change in attention to time cause by stimuli or environment). For simplicity, we ignore any other factors that can cause a change within the two components of the attentional-gate model.}
    \label{fig:attention_theory}
\end{figure}

The attentional-gate model \cite{zakay1995attentional} is a popular cognitive framework that explains time perception as an interplay between different components. As seen in Figure~\ref{fig:attention_theory}(a), it consists of a pacemaker that emits ticks at a given rate, referred to as clock speed. Clock speed can change due to different levels of arousal \cite{zakay1996role} caused by a stimuli or the environment. The ticks from the pacemaker pass through an attention-gate. When attention is focused on time, this gate is open wide and lets all the ticks pass through it. Conversely, when attention is diverted away from time, the gate is narrow and lets fewer ticks pass through it. The ticks that pass through the attention-gate are then accumulated in a cognitive counter. The reference memory contains information about the number of ticks needed in the cognitive counter to reach a target duration. The decision phase involves a comparator that compares the ticks in the cognitive counter with that in the reference memory, at different points in time. If the accumulated ticks are less than the reference memory ticks, then no action is taken. Otherwise, an action associated with marking the end of an interval may be taken. The clock speed, attention-gate width and reference memory depend on a number of factors, they differ between participants and can also change for a given participant in different situations.    

In this study, the attentional-gate model is used to formulate hypotheses about reasons for change in time perception. To simplify the problem, it is assumed that change in production time is a result of either a change in the cognitive counter or the reference memory or both (the two components immediately preceding the decision phase). Refer to Figure~\ref{fig:attention_theory}~(b) for a graphical representation. Using this simplification, we formulate a hypothesis about which features can indicate a change in a particular component of the attentional-gate model. Reference memory is affected by contextual calibrations like regression to the mean and prior timing experiences \cite{shi2013bayesian}. Hence, the time performance features, namely, prior-timing performance and self-evaluation of timing performance are assumed to indicate a change in reference memory. On the other hand, cognitive counter is affected by changes in attention to time or due to different levels of arousal caused by the stimuli or environment. Thus, the features under environmental characteristics and participant sensitivity are assumed to indicate a change in the cognitive counter. For simplicity, we ignore any other factors that can cause a change within the two components of the attentional-gate model.

Thus, the above assumptions enable aligning the selected features of the machine learning model with the different components of the attentional-gate model. Consequently, by combining these assumptions with feature importance analysis of the machine learning model, we can formulate a hypothesis about reasons for change in time production and provide a cognitive explanation  for the machine learning model's predictions. Specifically, when prior timing performance or self-evaluation of timing performance are found to have a significant contribution to the prediction, the change in time production is hypothesized to be caused by a change in reference memory of the participant. On the other hand, when environmental characteristics or participant sensitivity have a significant contribution to the prediction, the change in produced time is hypothesized to be caused by a change in the cognitive counter. It is important to note that the reasons for reference memory change, namely contextual changes, may not reflect an actual change in perceived time. Instead, it could be related to participants implicitly (or explicitly) adjusting their timing performance. Conversely, changes in cognitive counter can reflect an actual change in time perception because it involves changes in clock speed  or changes in attention to time.  In the Results section, we combine the discussions and assumptions made in this section with individual and population level feature importance results obtained using SHAP values, to provide a cognitive explanation for the  machine learning model's predictions.

\subsection{Baseline models}
\noindent The prediction model's performance is compared against two baseline models. These models are derived from hypotheses formulated using the attentional-gate model \cite{zakay1995attentional}.  
\begin{itemize}
    \item \textbf{Attention based model}: This model is based on the assumption that, a high engagement video distracts attention away from time towards the visual content. As a result, the cognitive counter would collect fewer ticks in a high engagement video trial as compared to a low engagement video trial. Hence, for the same objective time, the comparator would wait longer in case of a high engagement video as compared to a low engagement video, to match the reference memory ticks.  Thus, according to this model there would be an increase in production for a transition from lower to higher engagement level and conversely a decrease in time production for a transition from higher to lower engagement level.   

    \item \textbf{Arousal based model}: This model assumes that a high engagement video increases clock speed due to high arousal \cite{treisman1963temporal} as compared to a low engagement video. As a result, the cognitive counter would collect more ticks in a high engagement video trial as compared to a low engagement video trial. Here, we will see an opposite effect as compared to the attention based model described above. Thus, according to this model there would be a decrease in production for a transition from lower to higher engagement level and conversely a increase in time production for a transition from higher to lower engagement level.

\end{itemize}

\subsection{Second experiment to test generalization}
\noindent In order to check whether the model generalizes well to other scenarios, it was tested on data collected from an experiment with a slightly different setting. The details of this experiment are described in \cite{ugentStudyingHuman}. This was an offline experiment where participants were immersed in a driving simulator environment. The task was to drive through different scenes while maintaining a constant speed of 70 km/h and press a button when they thought 30 seconds had passed. After each trial, the participants were asked questions about the task difficulty, boredom and time perception, similar to the main experiment above. However, participants had to perform six trials in contrast to only two trials in the main experiment. Thus, while the second experiment was similar to the main experiment in terms of producing 30 seconds, it differed in terms of the non-timing task of maintaining a constant speed of 70 km/h (as opposed to watching the video with attention in the main experiment) and number of trials. Additionally, this was an offline setting while the main experiment was performed in an online setting.  

In order to adjust for six trials instead of two, we predict the time in the next trial given the information from the immediately previous trial. This additionally helped ensure that the model predictions are not limited to the first two trials but can generalize to any consecutive trials throughout the experiment. 
 
One more implementation difference between the two experiments was that in the second experiment, the different driving scenes did not have explicit labels of high, medium, or low engagement. Nevertheless, each scene projected a different level of difficulty in doing the non-timing task of maintaining a constant speed. This involved adding fog to the environment, a tail-gating car that could be seen from the rear view mirror, and a car in front that would break unexpectedly (stop-and-go). This difference was resolved by labelling a scene based on the performance in the non-timing task. The assumption was that if the non-timing task performance in a scene was poor on average, it was more engaging and diverted attention away from time. The stop-and-go scene had the worst performance and hence was labelled as a high engagement scene. In contrast, the experiment started with a scene called ``daylight", which did not involve stimuli or distractions, serving as a familiarization task for the participants. This was classified as a low engagement scene. All other cases, which fell between these two extremes in terms of performance, were classified as medium engagement scenes. Other features were derived from the data in the same way as in the last experiment. 

\section{Results}

\subsection{Direction of change prediction}

\noindent This section contains an evaluation of the logistic regression model for its ability to predict the direction of change in time production. When the predicted probability of decrease is more than 0.5, the prediction is classified as a decrease. Likewise, when the predicted probability is less than 0.5, the prediction is classified as an increase. 

The model was tested on both the main experiment data and data collected from the second experiment. For the main experiment, the model was evaluated using Leave-One-Out cross-validation (LOOCV). In this approach, the model is trained on all except one participant and tested on the left out participant. This process is repeated until each participant is considered as a test data point. Thus, the test data of the main experiment had 706 data points in total. For the second experiment, the test set contained 25 participants (29 participants performed the experiment, out of which, 4 were excluded for incomplete questionnaire). With each participant performing six trials, there were 125 data points in the second test set (last trial was excluded for all participants since there was no next trial to predict for).

Table \ref{table:performance} shows the precision (for decrease class), recall (for decrease class) and accuracy of the logistic regression model for the main experiment test set (Logistic Regression) and second experiment test set (Generalization results). The table also shows the results for the two baseline models (discussed in the Methods section) when tested on the data from the main experiment.  

The logistic regression model had an accuracy of 61\%. The precision of the model reveals that, out of all predictions suggesting a decrease in time production, 62\% were correct. Correspondingly, the recall reveals that, out of all participants that actually decreased their time production in the next trial, the model was able to predict 59\% of them correctly. Numerically the results may not seem very impressive. However, when compared with the rule-based baseline models that were derived from existing timing research, the model's accuracy is 10 percentage points higher. Furthermore, the model generalized fairly well on the second experiment data with an accuracy of 66\%. Since the two experiments had some crucial experimental differences (e.g. number of trials performed by a single participant), we should not compare the evaluation metrics for the two experiments. However, we can conclude that, on the second experiment data, the model performs better than a random model (which would have an accuracy of 50\% on a balanced dataset like the one used for model training and evaluation).   

Figure \ref{fig:prob_eval} shows the predicted probability for each participant. The correct and incorrect direction prediction can be derived from the figure as follows. All data points to the right of the vertical blue line (that indicates a probability of 0.5) are classified as decrease and those to the left are classified as increase by the model. Correspondingly, all points above the horizontal blue line (that indicates no change in produced time), represent an actual increase and those below represent an actual decrease. For the purpose of evaluating the direction of change prediction, the colour codes and grey dotted lines in the figure can be ignored. Thus, all points in the bottom-right and top-left quadrant formed by the two blue lines correspond to correct predictions. Whereas, all points in the top-right and bottom-left quadrant correspond to incorrect predictions.

\begin{table}
\begin{center}
{\renewcommand{\arraystretch}{1.5}%
{\begin{tabular}{  c c c c } 

\hline 
{\bfseries Model} & {\bfseries Precision} & {\bfseries Recall} & {\bfseries Accuracy} \\
\hline
Logistic Regression  & 0.617 & 0.589& 0.612 \\
\hline
Generalization results & 0.500 & 0.628 & 0.656 \\
\hline
Attention based model & 0.490 & 0.654 & 0.487 \\
\hline
Arousal based model & 0.519  & 0.346 & 0.513 \\

\hline
\end{tabular}}
\captionof{table}{Direction of change prediction: Model evaluation for it's ability to predict the direction of change in time production in the next trial. The table shows precision (for decrease), recall (for decrease) and accuracy of the prediction model on the main experiment test set (Logistic Regression), and second experiment test set (Generalization results). It also shows the results for the two baseline models (discussed in Methods) when tested on data from the main experiment.}
\label{table:performance}
}
\end{center}
\end{table}

\subsection{Magnitude of change}

\noindent This section contains an evaluation of the logistic regression model's output probabilities to determine the amount of information they carry about magnitude of change in time production. For simplicity, the magnitude of change is categorized into three levels: ``high decrease", ``high increase" and ``small change". We assume that for highly informative probabilities, a high probability of decrease should correspond to a high decrease in production time, a low probability of decrease should correspond to a high increase in production time, and a probability close to 0.5 which indicates uncertainty in model prediction should correspond to a small change.

\begin{figure}
    \centering
        \includegraphics[width=\linewidth]{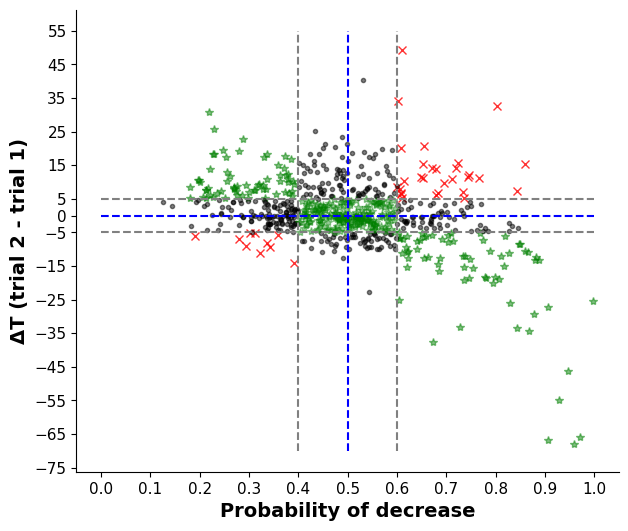}

    \caption{Probability evaluation for magnitude and direction of change prediction: The graph shows the trained logistic regression model's predicted probability of decrease against $\Delta$T (difference in time produced between two trials) for all participants (706). The probability of decrease was obtained using LOOCV. These probabilities have been divided into three levels (indicated by the vertical grey dotted lines): low (probability $<$ 0.4), uncertain (0.4 $<$ probability $<$ 0.6) and high (probability $>$ 0.6). Similarly, $\Delta$T has been divided into three levels (indicated by the horizontal grey dotted lines): high decrease ($\Delta$T $<$ -5), small change (-5 $<$ $\Delta$T $<$ 5) and high increase ($\Delta$T $>$ 5). The green *'s correspond to the cases when the predicted probability conveyed the correct information about magnitude of change level, the red X's correspond to cases when the predicted probability conveyed extremely incorrect information, and the black dots correspond to other slightly incorrect cases. Table \ref{table:cm_magnitude} shows the number of data points in each of the nine regions created by the grey dotted lines (both horizontal and vertical).
    The vertical and horizontal blue lines corresponds to probability of 0.5 and $\Delta$T of zero respectively. The four quadrants formed by these lines can be used as reference to evaluate the direction of change predictions. All points in the bottom-right and top-left correspond to correct direction of change prediction and all points to the top-right and bottom-left correspond to incorrect predictions.
    }
    \label{fig:prob_eval}
\end{figure}

The difference in time produced between two trials is referred as $\Delta$T (Next trial production time - Previous trial production time). We define ``small change" in production time as corresponding to $\Delta$T values between -5 and 5 (indicated by the grey horizontal dotted lines in Figure \ref{fig:prob_eval}).  Consequently, ``high increase" corresponds to $\Delta$T $>$ 5, where the next trial's production time exceeds the previous trial's production time by more than 5 seconds. Similarity, a ``high decrease" corresponds to $\Delta$T $<$ 5, where the next trial's production time is less than the previous trial's production time by more than 5 seconds. We define probabilities between 0.4 and 0.6 as corresponding to uncertainty in model predictions (indicated by the grey vertical dotted lines in figure \ref{fig:prob_eval}). Consequently, high probability of decrease corresponds to output probabilities more than 0.6 and low probability of decrease (or high probability of increase) corresponds to output probabilities below 0.4. The choice of uncertainty range for probabilities and the limits for defining small $\Delta$T are based on heuristics and solely serve the purpose of simplifying the analysis. While changing these ranges or limits will not affect the model's output probabilities, it will alter the aggregated results presented in this section.  Thus, the results discussed in this section, only aim to build an intuition about the amount of information contained in the logistic regression model's output probabilities. 

For simplicity, we only consider five extreme cases for evaluation of model's ability to predict magnitude of change. First, when model predicts high increase (probability $<$ 0.4) and there is actually a high increase ($\Delta$T $>$ 5). Thus, the model probabilities in this case convey correct information about magnitude of change (green *'s in top left of Figure \ref{fig:prob_eval}). Second, when model predicts high increase (probability $<$ 0.4) but there is instead a high decrease ($\Delta$T $<$ -5) in time production. Thus, the model probabilities in this case convey extremely incorrect information about magnitude of change (red X's in bottom left of Figure \ref{fig:prob_eval}). Third, when model predicts a small change (probability between 0.4 and 0.6) and the actual change in time production is also small ($\Delta$T between -5 and +5). Thus, model probability conveys the correct information about uncertainty or a small change (green *'s in the centre of Figure \ref{fig:prob_eval} ). Fourth and fifth cases correspond respectively to first and second, but when predicted probability is high (more than 0.6). All other cases (black dots in Figure \ref{fig:prob_eval}), are not considered. They correspond to cases when the model probabilities convey incorrect information but with a small margin (a difference of one level of magnitude between actual and predicted magnitude level). It is important to note that, although they correspond to slightly incorrect magnitude information, these cases include both correct and incorrect direction predictions.  For example, when model predicts a high decrease and if the actual magnitude of change is small, the change may either be a small decrease (correct direction prediction) or a small increase (incorrect direction prediction).

Table \ref{table:cm_magnitude} shows the confusion matrix for the three magnitude of change levels. The coloured numbers correspond to the five extreme cases discussed above. Thus, we can conclude that, when the model's output probability suggests a high increase (predicted probability $<$ 0.4), it is correct 34\% (65/189)  of the times and extremely incorrect 5\% (10/189) of the times. Similarly, when the model's output probability suggests a high decrease (predicted probability $>$ 0.6),  it is correct 42\% (68/162) of the times and extremely incorrect 18\% (29/162) of the times. Additionally, when the model's output probability suggests a small change in time production (0.4 $<$predicted probability $<$ 0.6), it is correct 57\% (202/355) of the times. 

In order to test how well the model generalizes for magnitude prediction, the same analysis is performed on the second experiment data. Table \ref{table:cm_magnitude_driving_sim} shows the results. The model is correct 42\% (19/45)  of the times and never extremely incorrect in suggesting a high increase. Similarly, it is correct 33\% (13/39) of the times and extremely incorrect 5\% (2/39) of the times in suggesting a high decrease. And, it is correct 78\% (32/41) of times in suggesting a small change. 

Thus, the probabilities of the trained logistic regression model convey additional information which a rule-based model cannot achieve. This also means that we do not have any baselines (derived from timing research) to compare these results against. Nevertheless, by looking at Figure \ref{fig:prob_eval} and tables \ref{table:cm_magnitude} and \ref{table:cm_magnitude_driving_sim}, the model probabilities can be considered reliable because of low numbers of extreme incorrect cases. The results are promising also because the model was not explicitly trained for the three magnitude levels. Instead, it was only trained on the binary task of predicting the direction of change. The resulting probabilities happen to carry some information about the magnitude of change. Thus, an approximate inference about how long a participant will wait before stopping the video can be made using the probabilities of the logistic regression model.

\begin{table}
\begin{center}
{\renewcommand{\arraystretch}{1.5}%
{\begin{tabular}{  c c c c } 

\hline 
\diagbox[width=60pt]{Actual}{Predicted} & {\bfseries High increase} & {\bfseries Small change} & {\bfseries High decrease}   \\
\hline
\textbf{High increase}  & \textcolor{PineGreen}{65 (34\%)} & 95 & \textcolor{red}{29 (18\%)}\\
\hline
\textbf{Small change} & 114 & \textcolor{PineGreen}{202 (57\%)} & 65 \\
\hline
\textbf{High decrease} & \textcolor{red}{10 (5\%)} & 58 & \textcolor{PineGreen}{68 (42\%)} \\
\hline
Total & 189 & 355 & 162\\

\hline
\end{tabular}}
\captionof{table}{Confusion matrix for magnitude of change prediction on main experiment data. The numbers correspond to the number of data points in each of the nine regions formed due to the grey dotted lines (horizontal and vertical) in Figure \ref{fig:prob_eval}. The evaluation focuses on the five extreme cases discussed in the magnitude of change analysis (represented in green and red).}
\label{table:cm_magnitude}
}
\end{center}
\end{table}

\begin{table}
\begin{center}
{\renewcommand{\arraystretch}{1.5}%
{\begin{tabular}{  c c c c }

\hline 
\diagbox[width=60pt]{Actual}{Predicted} & {\bfseries High increase} & {\bfseries Small change} & {\bfseries High decrease}   \\
\hline
\textbf{High increase}  & \textcolor{PineGreen}{19 (42\%)} & 7 & \textcolor{red}{2 (5\%)} \\
\hline
\textbf{Small change} & 26 & \textcolor{PineGreen}{32 (78\%)} & 24 \\
\hline
\textbf{High decrease} & \textcolor{red}{0} & 2 & \textcolor{PineGreen}{13 (33\%)} \\
\hline
Total & 45 & 41 & 39\\

\hline
\end{tabular}}
\captionof{table}{Confusion matrix for magnitude of change prediction on second experiment data.}
\label{table:cm_magnitude_driving_sim}
}
\end{center}
\end{table}

\subsection{Cognitive interpretation of the prediction model}

\noindent This section highlights the most important features and illustrates how the different feature values affect the model's predictions at a population level. We first explain the working of the prediction model at a population level using aggregated SHAP values \cite{scott2017unified} and the logistic regression model's coefficients from Equation \ref{eq:X_trained}. We then add a cognitive explanation to the results of this analysis using the feature alignment assumptions from Figure \ref{fig:attention_theory} (b).  

\begin{figure}[!t]
    \centering
    \includegraphics[width=\linewidth]{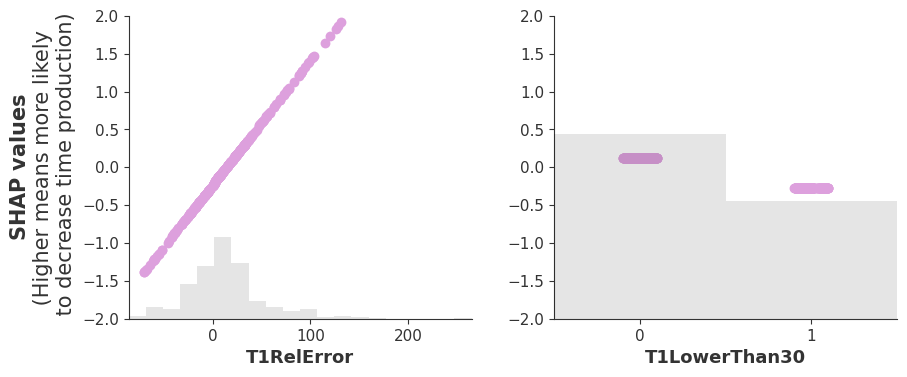}
    \includegraphics[width=\linewidth]{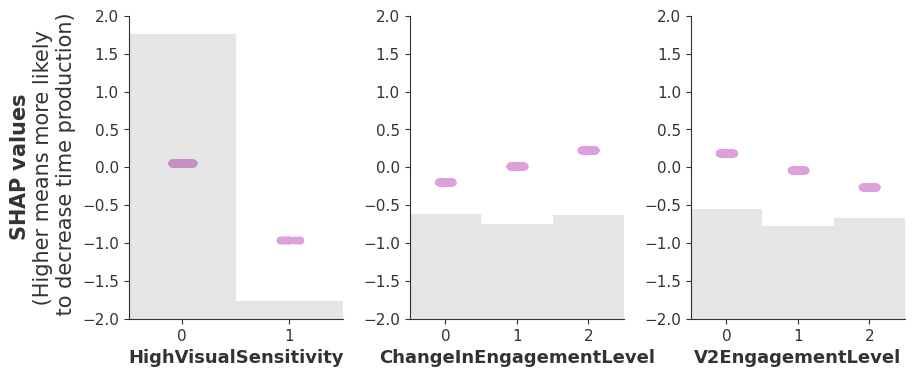}
    \caption{Cognitive interpretation of the prediction model: Each plot shows the aggregated SHAP values (shown as pink dots) for different features. Since SHAP values correspond to feature contribution, very high and very low values indicate larger impact on prediction. Large positive values indicate a positive impact on prediction (supports probability of decrease in this case) and large negative values indicate a negative impact on prediction (supports probability of increase in this case). Consequently, SHAP values close to zero indicate smaller impact on prediction. The grey histograms in each plot show the distribution of the respective feature values. }
    \label{fig:global_analysis: shap}
\end{figure}

The prior timing feature, \textit{T1RelError} is positively correlated with the probability of decrease. This can be inferred from the aggregated SHAP values of this feature (Figure \ref{fig:global_analysis: shap}, top-left plot). Extreme positive values of this feature correspond to high positive SHAP values. This indicates a significant contribution of this feature towards increasing the predicted probability, resulting in a high probability of decrease. Conversely, extreme negative values correspond to high negative SHAP values. This indicates a significant contribution of the feature in decreasing the predicted probability, resulting in a very low probability of decrease (or a high probability of increase).  As the feature value moves closer to zero from the extremes, the SHAP values also move closer to zero indicating a reduced impact of this feature on the predicted probability. An intuitive explanation of this analysis is that, when trial 1 production time is significantly higher than target 30 seconds (i.e. T1RelError takes high positive values), the predicted probability of decrease in the next trial is very high. Conversely,  when trial 1 production time is significantly lower than 30 seconds (i.e. T1RelError takes high negative values), the predicted probability of increase is very high. The point at which the contribution of this feature switches from increase to decrease or vice versa is when trial 1 production time is equal to its mean value of 34 seconds (refer Supplementary information for data analysis results). According to our attentional-gate theory assumption, this feature is related to the reference memory (Figure \ref{fig:attention_theory} (b)). Thus by combining the SHAP value analysis with the attentional-gate theory, we can formulate two hypothesize. First, participants explicitly calibrate their reference memory in the subsequent trial, to adjust their production time towards the target. Hence, the model captures participants' tendency to correct their timing performance in the next trial. Second, participants may implicitly calibrate their reference memory to adjust their production time towards the mean. In this case, the model captures the regression to the mean phenomenon \cite{barnett2005regression}, which is a well-documented occurrence in timing research \cite{lejeune2009vierordt}.

The self-evaluation of timing performance feature, \textit{T1LowerThan30}, is negatively correlated to the probability of decrease. This can be inferred from the aggregated SHAP values (in Figure \ref{fig:global_analysis: shap}, top-right plot). Intuitively, this means that, if participants report being higher than 30 seconds in trial 1, their probability of decrease in the next trial is higher. Likewise, when participants report being lower than 30 seconds, their probability of increase is higher. According to our attentional-gate theory assumption (Figure \ref{fig:attention_theory}(b)), this feature is again related to changes in reference memory. Thus, this analysis supports the hypothesis that participants actively correct their timing (i.e. explicitly calibrate their reference memory) based on their explicit evaluation of previous timing performance.  

The aggregated SHAP values of the participant sensitivity feature, \textit{HighVisualSensitivity}, show that participants who are marked as sensitive have a very high probability of increase (Figure \ref{fig:global_analysis: shap}, bottom-left). When this feature is True, the corresponding SHAP value is very low (-1.0) in comparison to other features, indicating a high impact of this feature on final prediction. However, when the feature is not set to True, which is the case for most (94\%) participants, the SHAP values indicate minimal impact on the probability of decrease. Nevertheless, 68\% of participants marked as sensitive by this feature increase their time in the next trial. Thus, this feature being True is a good indicator of increase in time production. According to Figure \ref{fig:attention_theory}(b), this feature is related to changes in the cognitive counter. Thus, we hypothesize that participants marked as sensitive are more likely to experience distractions and shift focus away from time, resulting in a narrower attention-gate width in the next trail. Due to this, the cognitive counter accumulates fewer ticks, resulting in an increased production time. Hence, the model identifies these participants as being more prone to increasing their time production irrespective of the video engagement level in the subsequent trial. 

The feature, \textit{V2EngagementLevel}, which corresponds to the video engagement level in trial 2, shows a negative correlation with the probability of decrease (refer Figure \ref{fig:global_analysis: shap}, bottom-right plot). Intuitively, this means that a low engagement video in trial 2 contributes to a high probability of decrease, while a high engagement video contributes to a high probability of increase. According to our attentional-gate model assumption, this feature which comes under environmental characteristics, is related to changes in cognitive counter (Figure \ref{fig:attention_theory}(b)). Based on the analysis, we hypothesize that the engagement level of trial 2 video is related to the attention-gate width. A high engagement video in trial 2 draws attention away from time, narrowing the attention-gate width. Due to this, the cognitive counter accumulates fewer ticks resulting in an increased production time in trial 2.  Conversely, a low engagement video in trial 2 has an opposite effect, widening the attention-gate width. This causes more tick accumulation in the cognitive counter, resulting in a decreased production time in trial 2.

The second environmental characteristic feature, \textit{ChangeInEngagementLevel}, shows a positive correlation with the probability of decrease (Figure \ref{fig:global_analysis: shap}, bottom-centre). This feature captures the change in engagement level between trial 1 and trial 2 video. Intuitively, this means that going from a lower engagement video in trial 1 to a higher engagement video in trial 2 (\textit{ChangeInEngagementLevel} =2 in Figure \ref{fig:global_analysis: shap}), contributes towards a higher probability of decrease. Conversely, going from a higher engagement to a lower engagement video (\textit{ChangeInEngagementLevel}=0 in Figure \ref{fig:global_analysis: shap}), contributes towards a higher probability of increase. The SHAP values for cases when there is no change in engagement level for the two consecutive videos (\textit{ChangeInEngagementLevel} =1 in Figure \ref{fig:global_analysis: shap}), indicate no impact of this feature on the prediction probability.  Combining this analysis with our attentional-gate model assumption (Figure \ref{fig:attention_theory}), we hypothesize that this feature is related to arousal and thus changes in clock speed. A shift from lower engagement to a higher engagement video increases arousal and thus clock speed. This causes more tick accumulation in the cognitive counter, resulting in a decreased time production in trial 2. Similarly, a shift from higher engagement to a lower engagement video reduces arousal and thus clock speed, causing fewer tick accumulation in the cognitive counter. Thus resulting in an increased time production in trial 2.

Based on the coefficients in Equation \ref{eq:X_trained}, we can conclude that the prior timing feature has a significant impact on model predictions at a population level. The aggregated SHAP values (in Figure \ref{fig:global_analysis: shap}) provide further insights into this observation. They reveal that, extreme values of the prior timing feature, as well as the participant sensitivity feature when marked as True, significantly affects the predicted probability of the logistic regression model.

\subsection{Cognitive interpretation of individual predictions}

\begin{figure}[!t]
    \centering

    \includegraphics[width=\linewidth]{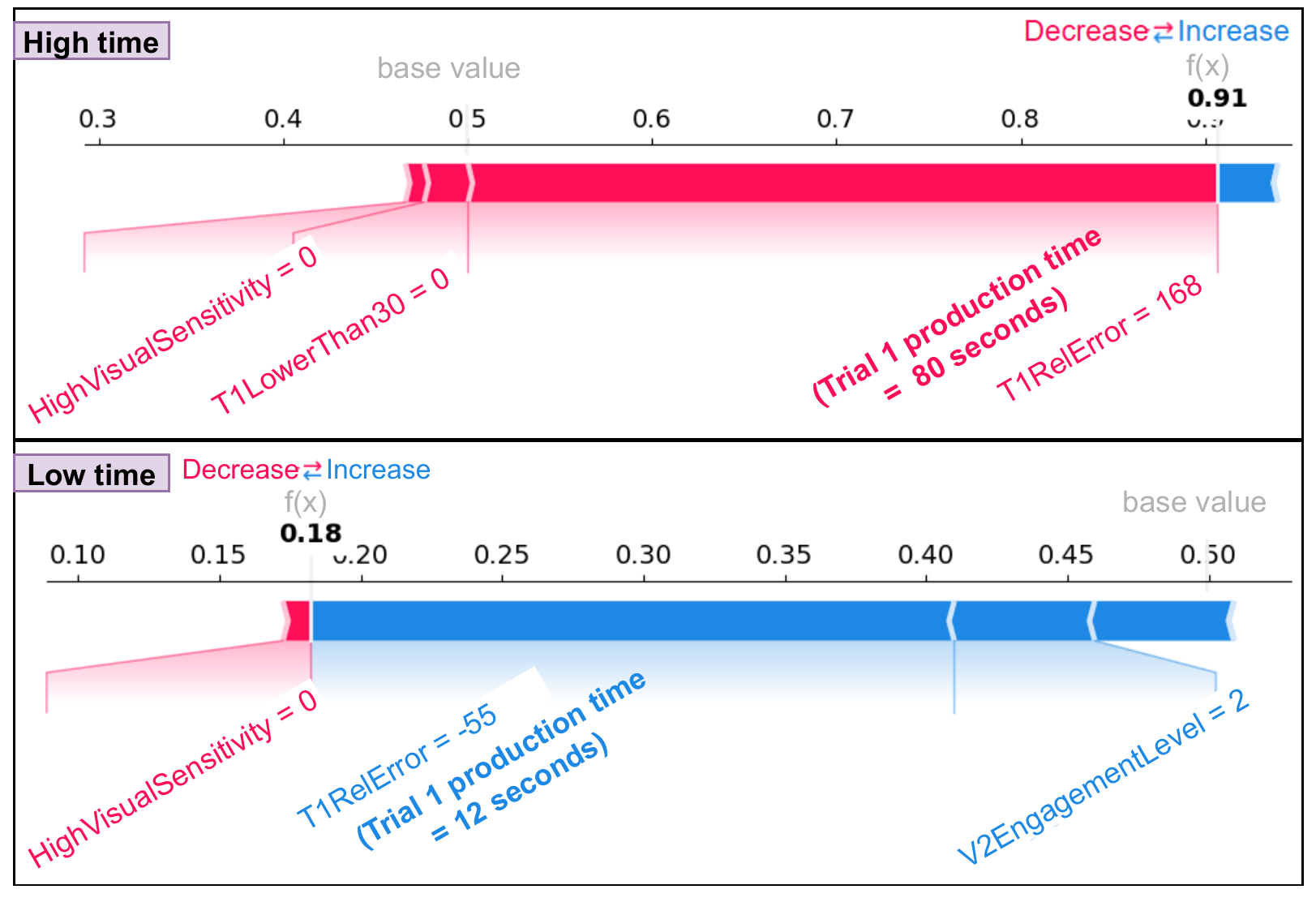}
    \caption{Example of feature contributions for participants with extreme values of trial 1 production time. Features shown in red (pointing to the right) move the the predicted probability to higher values. These feature contribute towards suggesting a decrease in time production. Likewise, features shown in blue (pointing to left) move the predicted probability to lower values and hence contribute towards suggesting an increase in time production. The top plot shows an example when trial 1 production time is 80 seconds and hence \textit{T1RelError} is 168\%. The bottom plot shows an example when trial 1 production time is 12 seconds and hence \textit{T1RelError} is -55\%. In both cases, the prior timing performance feature, \textit{T1RelError}, has the maximum impact on prediction in comparison to other features. It has a significant contribution in moving the predicted probability from it's base value of 0.5 towards the extremes.}
    \label{fig:shap_local_time}

\end{figure}

\begin{figure}
    \centering
   \includegraphics[width=\linewidth, height=10cm]{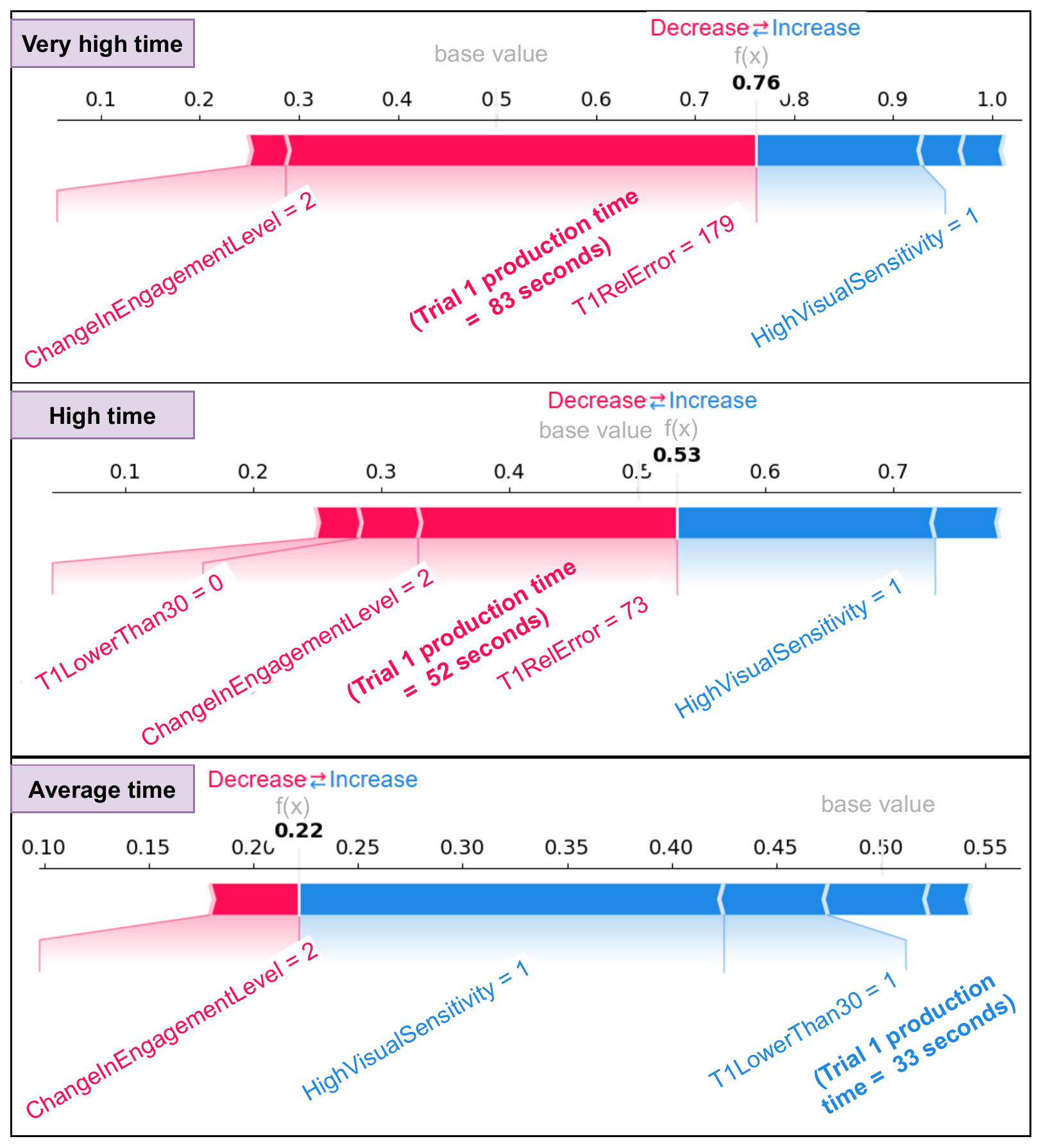}
    \caption{Example of feature contributions for three levels of prior timing performance, when participant sensitivity (\textit{HighVisualSensitivity}) is marked as True. The top image corresponds to trial 1 production time of 83 seconds (or \textit{T1RelError} = 179\%). Similarly, the middle image corresponds to trial 1 production time of 52 seconds (or \textit{T1RelError} = 73\%) and bottom to trial 1 production time of 33 seconds or \textit{T1RelError} of 10\% (not seen in the image). In all cases, the participant sensitivity feature shifts the predicted probability to suggest an increase in time production, with varying levels of impact. As trial 1 production time moves closer to it's mean, the impact of participant sensitivity feature on predicted probability increases.}
    \label{fig:shap_local_sensitivity}
\end{figure}

\begin{figure}[!t]
    \centering
    \includegraphics[width=\linewidth]{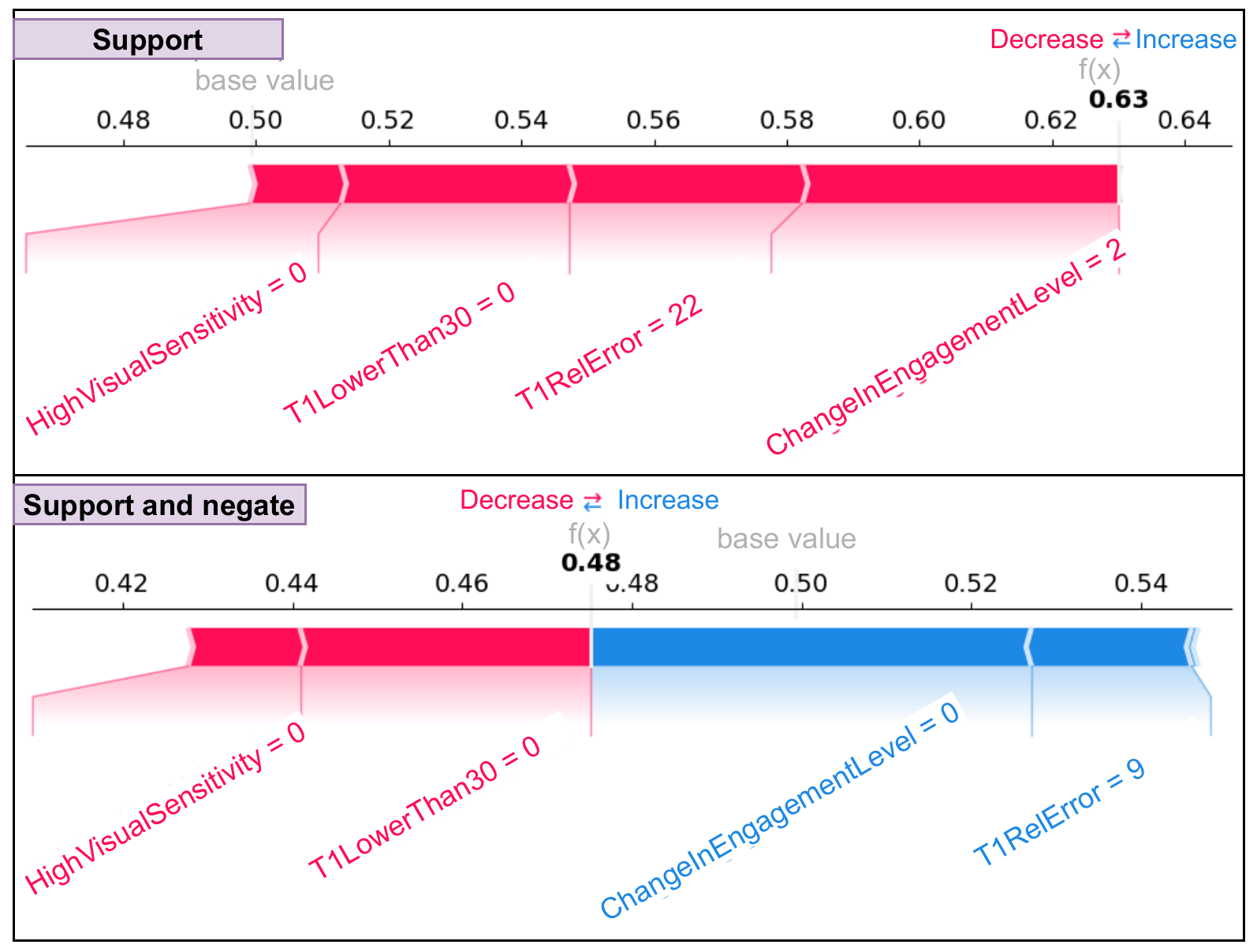}
    \caption{Example of feature contributions when no single feature has a significant impact on prediction. The top image shows an example when all features support each other. The resulting probability is a cumulative effect of small contributions from all the features and tends to be moderately high (0.63). The bottom image shows an example of features negating each other. The resulting probability (0.48) thus remains close to the base value of 0.5, indicating uncertainty in model prediction. }
    \label{fig:shap_local_others}
\end{figure}

\noindent This section shows some examples for analysing the combined effect of different feature values on a specific prediction (individual level analysis). It gives insights into how the model makes predictions for a specific participant. We integrate this analysis with the feature alignment assumptions from Figure \ref{fig:attention_theory}(b) to formulate hypotheses about reasons for change in a participant's time production. The analysis is carried out using SHAP values for individual predictions. All examples used in this analysis are selected from the test set and correspond to correct predictions by the logistic regression model.  

Since the prior timing feature had a significant impact on model predictions for extreme values, we consider one example each of very high (Figure \ref{fig:shap_local_time}, top image) and very low (Figure \ref{fig:shap_local_time}, bottom image) values of the feature. In both cases, the prior timing performance feature has a significant contribution in moving the predicted probability from its base value of 0.5 towards the extremes. Other features have a very small contribution in influencing the final predicted probability. Thus, the model majorly basis it's prediction on the prior timing feature when it takes on extreme values. This results in either a very high output probability of decrease (0.91, see Figure \ref{fig:shap_local_time}, bottom image) or very high probability of  increase (indicated by a low output probability of decrease, 0.18 in Figure \ref{fig:shap_local_time}, bottom image).  According to our attentional-gate theory assumption (Figure \ref{fig:attention_theory} (b)), these participants majorly change production time due to changes in reference memory instead of the cognitive counter. This assumption in turn implies that the change in production time for these participants may not reflect an actual shift in their time perception, which is typically linked to changes in cognitive mechanisms like attention-gate width or pacemaker clock speed.

The feature related to participant sensitivity when True, also had a significant impact on model predictions. Figure \ref{fig:shap_local_sensitivity} shows three examples to analyse the interplay between this feature value and three levels of the prior timing performance feature values. For extremely high values of the prior timing performance feature (top image in Figure \ref{fig:shap_local_sensitivity}), the participant sensitivity feature has a relatively small impact on the final prediction. It shifts the final probability only slightly towards a lower value or towards increase (indicated in blue in Figure \ref{fig:shap_local_sensitivity}). The same effect is observed for very low values of the prior timing performance feature (not shown in the Figure). However, as the prior timing performance moves closer to the mean (centre and bottom image in Figure \ref{fig:shap_local_sensitivity}), the impact of participant sensitivity feature increases, and that of prior timing performance decreases. Based on our attentional-gate theory assumption (Figure \ref{fig:attention_theory}(b)), we hypothesize that when a participant marked as sensitive, produces a trial 1 time close to the population mean, the change in time production is due to a change in the cognitive counter instead of the reference memory. This further implies that, for such participants, the change in time production can reflect an actual shift in time perception. This shift is hypothesized to result from attention being diverted away from time, as discussed in the cognitive interpretation of the prediction model analysis. 

When prior timing performance feature value is close to its mean, and participant sensitivity is not marked as True, the final probability is influenced almost equally by all the features. Since the feature contributions in this case are small, the output probability is moderately high only when all features support each other. Otherwise, when the features tend to negate each other, the output probability remains close to 0.5 indicating uncertainty in model prediction. The top plot in Figure \ref{fig:shap_local_others} shows an example when all features support each other and shift the output probability from it's base value of 0.5 towards a higher probability of decrease (0.63). Conversely, the bottom plots in Figure \ref{fig:shap_local_others} shows an example when the features tend to negate each other. In this case, the resulting probability (0.48) remains close to the base value of 0.5, indicating uncertainty in model prediction. Based on our attentional-gate theory assumption (Figure \ref{fig:attention_theory}(b)), we hypothesize that the change in time production for these participants is due to changes in both the reference memory and the cognitive counter.  

The aggregated SHAP values at a sub-group level for participants with similar feature value combinations reveal similar interactions between the features. For further details on sub-group level analysis, refer to the supplementary information.

\section{Discussion}

\noindent Through this study we attempted to move closer to a quantitative analysis of time perception using techniques from machine learning on an ecologically valid dataset. Using features like previous timing performance, explicit self-evaluation of timing performance, participant sensitivity, and, environmental characteristics (video engagement levels), the machine learning model predicts the probability of decrease. The accuracy of this model (61~\%) was better than two rule-based baseline models derived from timing research. In addition to direction of change, the model's output probabilities also carried information about the magnitude of change. The machine learning model was further tested on data from a second experiment where the accuracy (66\%) was better than a random model. Thus, the model exhibits some generalization capabilities. The generalization results further showed that the model can be used for any consecutive trials in a sequence of trials and does not have to be limited to the first two trials. This suggests that timing performance in a given trial is related to some characteristics of the immediately preceding trial.

In addition to training a prediction model, this study also attempts to explain the model's predictions using SHAP values, a technique from Explainable AI. One significant and consistent finding from the population and individual level feature analysis was that previous timing performance plays a significant role in determining the direction of change in time production for the upcoming trial. We formulate two hypotheses based on this finding. First, participants regress towards the mean. As noted in the feature analysis of prior timing performance, the model learns to use the mean of trial 1 production time as a threshold to change the direction of impact of this feature on the final probability. Regression to the mean is a common phenomena in time perception research \cite{lejeune2009vierordt} and hence can be considered a plausible explanation for the model's behaviour. Second, participants tend to adjust their timing in the subsequent trial so as to be closer to the target interval. The feature analysis results of the self-evaluation of timing performance feature supports this hypothesis. This is because, the model identifies a pattern where participants who classify their production time as being higher than the target, tend to reduce their production time in the subsequent trial and vice versa. We do not have enough evidence in this study to prove or disprove either of the two hypotheses.

One potential application of this model is in systems like ChronoPilot \cite{botev2021chronopilot}. In this system, the goal is to modulate the perceived time of a user performing a task within a VR/AR environment, in order to promote well-being and better task performance. We envision that the probabilities output by the model can be used to make a decision about the next action to take in order to modulate time in the desired direction. For example, if the model outputs a high probability of decrease and if the final goal is to increase time production (which indicates higher attention to a non-timing task), we could activate certain modulators (based on timing research) that draw attention away from time towards the non-timing task. Alternatively, if the model outputs a very low probability of decrease, no action is needed by the system. An alternative interpretation of the model's prediction in such systems could be that a very high probability of decrease indicates that it is difficult to modulate time such that the participant will experience an increase in time perception. In such cases, we can make decisions to use more powerful modulators or do nothing. Additionally, the feature analysis performed in this study facilitates understanding the contribution of each feature to a specific prediction. This information can further aide the decision making process in systems like ChronoPilot.  

By combining the attentional-gate model (Figure \ref{fig:attention_theory}) of time perception with the population and individual level feature analysis performed in this study, we can formulate hypotheses about the reasons for change in production time for a specific participant. Additionally, this enables us to hypothesize whether the change in production time reflects an actual shift in time perception. For example, when model predicts a decrease with a high probability and the corresponding value of the prior timing performance feature is very high, it can imply that the participant will decrease time due to changes in reference memory (caused either by regression to the mean phenomena or active time correction by the participant). This further suggests that the change in time production is not a result of an actual shift in time perception, which is typically related to changes in clock speed or attention to time. On the other hand, when the model predicts a decrease with a high probability, and the features with a significant impact on prediction are related to the cognitive counter instead of the reference memory, it implies that there was either a change in clock speed or attention to time. This suggests that the decrease in time production in this case was due to an actual shift in time perception between the two trials. Knowing whether or not participants will actually experience time to go faster/slower can be very useful in making decisions that involve taking actions based on a participant's perception of time. 

The analysis conducted in this study has a few limitations. Based on the analysis, we cannot confirm the different hypotheses discussed using the attentional-gate model theory about reasons for change in time production. We can only quantitatively analyse some correlations between the different controllable features (like engagement level of the upcoming video), uncontrollable features (like prior timing performance and participant sensitivity), and the final model predictions. Thus, while we can determine the feature contributions for a given prediction, we cannot directly confirm our attentional-gate theory assumptions (Figure \ref{fig:attention_theory}(b)) about the causal relationships between the features and direction of change in time production. Whether the features used by the machine learning model are related to the reference memory or the cognitive counter, and whether a change in time production represents an actual shift in time perception also requires further investigation.

One potential future work for this study is performing a causal analysis to investigate whether the controllable features used by the machine learning model influence the direction of change in time production in the subsequent trial. Such an analysis can provide insights into whether changing features like video engagement level (which can be controlled by the experimenter) of the upcoming trial can cause a change in production time of an individual. The results of such a study can be considered a small step towards influencing the time produced by an individual in a subsequent trial. Furthermore, complementing this study with an analysis of whether a change in time production reflects an actual change in time perception can provide valuable insights for modulating the time perception of an individual. 

\section{Acknowledgments}
\noindent This work was supported by European Union’s Horizon 2020 FET research program under grant agreement
No. 964464 (ChronoPilot). We thank Argiro Vatakis and Eirini Balta for their assistance in designing the main experiment used for data collection. 

\bibliographystyle{unsrt}
\bibliography{references}

\newpage
\textbf{\huge Appendix}
\onecolumn

\section{Supplementary information}

\subsection{Data analysis}
Total participants: 986

\subsubsection{Production time in the two trials was significantly different than target interval (30 seconds)}

\begin{itemize}
    \item Trial 1 (M=33.8, STD=12.4) and Trial 2 (M=36.5, STD=12.7) time is significantly different than 30 seconds (p $\ll$ 0.05).
    \item Z-test statistics trail 1 - 9.59, p $\ll$ 0.05
    \item Z-test statistics trail 2 - 15.99, p $\ll$ 0.05

\end{itemize}

\begin{figure}
    \centering
    \includegraphics[width=0.8\linewidth, height=10cm]{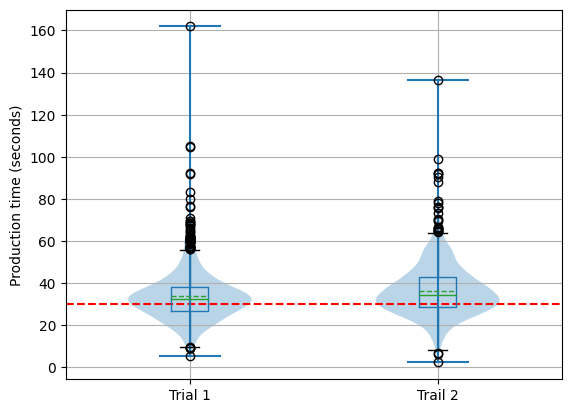}
    \caption{Production time for trial 1 and trial 2. The red dotted line represents objective 30 seconds. The dotted green line in the box plots show the means and solid green line represents the median of each distribution.}
    \label{fig:data_analysis_raw_production_time}
\end{figure}

\noindent Figure \ref{fig:data_analysis_raw_production_time} shows the distribution of trial 1 and trial 2 production times. Table below shows the statistics for each trial's production time.
\\
\begin{center}
\begin{tabular}{c  m{8em} m{8em}  } 

\hline 
{} & {Trial 1 production time} & { Trial 2 production time}   \\
\hline
count &                 986.0 &                 986.0 \\
\hline
mean  &                  33.8 &                  36.5 \\
\hline
std   &                  12.4 &                  12.7 \\
\hline
min   &                   5.2 &                   2.3 \\
\hline
25\%   &                  26.8 &                  28.4 \\
\hline
50\%   &                  32.5 &                  34.6 \\
\hline
75\%   &                  38.3 &                  42.7 \\
\hline
max   &                 161.9 &                 136.5 \\
\hline
\end{tabular}
\end{center}

\subsubsection{Trial 1: Difference between  time produced for different video engagement levels in trial 1}

\begin{itemize}
    \item Low engagement not significantly different than medium engagement. (t-test statistic=0.34, pvalue=0.73) 
    \item Low engagement not significantly different than high engagement. (t-test statistic=0.34, pvalue=0.73) 
    \item Medium engagement not significantly different than high engagement. (t-test statistic=-0.001, pvalue=0.99)  
\end{itemize}

\noindent Table below shows the statistics for the production times of different video engagement levels in trial 1 \\

\begin{center}
\begin{tabular}{c m{8em} m{6em} m{8em} }
\hline
{} &  Low engagement&  Medium engagement &  High engagement \\
\hline
count &           331.0 &              325.0 &            330.0 \\
mean  &            34.0 &               33.7 &             33.7 \\
std   &            11.5 &               12.8 &             12.8 \\
min   &             9.4 &                9.0 &              5.2 \\
25\%   &            27.4 &               26.6 &             26.3 \\
50\%   &            32.7 &               32.6 &             32.0 \\
75\%   &            38.3 &               38.4 &             38.3 \\
max   &           105.2 &              161.9 &            104.6 \\
\hline
\end{tabular} 
\end{center}

\subsubsection{Trial 2: Difference between time produced for different video engagement levels in trial 2}
\begin{itemize}
    \item Low engagement not significantly different than medium engagement time. (statistic=-1.83, pvalue=0.06) 
    \item Low engagement \textbf{significantly} different than high engagement. (t-test statistic=-2.44, pvalue=0.01)
    \item Medium engagement not significantly different than high engagement. (t-test statistic=-0.75, pvalue=0.45) 
\end{itemize}

\noindent Table below shows the statistics for the production times of different video engagement levels in trial 1 \\

\begin{center}
    
\begin{tabular}{c m{6em} m{6em} m{6em} }
\hline
{} &  Low engagement &  Medium engagement &  High engagement \\
\hline
count &           326.0 &              340.0 &            320.0 \\
mean  &            35.1 &               36.8 &             37.5 \\
std   &            12.2 &               12.2 &             13.6 \\
min   &             2.3 &                9.7 &             10.4 \\
25\%   &            28.0 &               29.0 &             28.8 \\
50\%   &            33.0 &               35.7 &             34.9 \\
75\%   &            40.0 &               43.4 &             43.3 \\
max   &           136.5 &               88.1 &             99.0 \\
\hline
\end{tabular}
\end{center}

\subsubsection{Comparing time production in the two trials}
There was an increase in the average time in trial 2 (M=36.5, STD=12.) as compared to trial 1 (M=33.8, STD=12.4). The average production time in trail 2 is significantly more than that in trial 1. (related t test statistics=8.36, pvalue $\ll$ 0.05). This could mean that attention was diverted away from time in trial 2 as compared to trial 1 (irrespective of the video engagement level in the two trials).\\ 

\subsubsection{Shift in attention}

\begin{figure}
    \centering
    \includegraphics[width=0.8\linewidth, height=7cm]{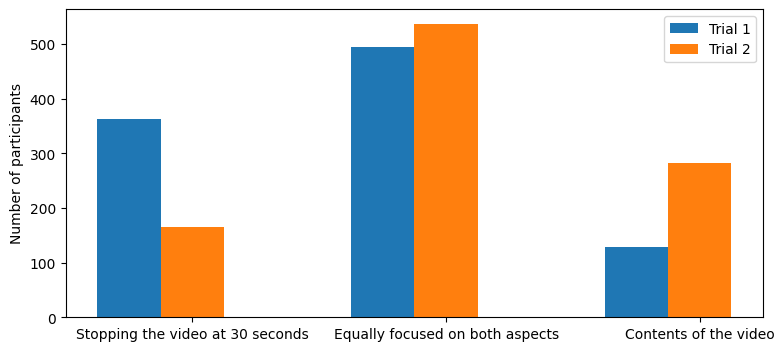}
    \caption{Comparing the number of participants attending to timing or non-timing task in the two trials}
    \label{fig:data_analysis_attention}
\end{figure}

In the questionnaire following the videos in each trial, participants were asked whether they focused on stopping the video at 30 seconds or the contents of the video or both. Considering this question to indicate whether the participant's attention was directed towards time (reported to focus on timing task) or away from time (reported to focus on non-timing task), Figure \ref{fig:data_analysis_attention} shows shift in attention to time between the two trials. The number of participants who report to focus on time (stopping the video at 30 seconds) is higher in trial 1 as compared to trial 2. Conversely, the number of participants reporting to focus on non-timing task (Content of the video) is lower in trial 1 as compared to trial 2. One hypothesis for this shift in attention is that the questionnaire is introduced for the first time only after the trial 1 video. This questionnaire contains questions about video content. Hence, in the trial 1 video, most participants focus on the timing task. Once introduced to the kind of questions that will be asked about video content, the number of participants focusing on video content increases. \\

\subsubsection{Change in time production between trials}

The table below shows some statistics for $\Delta$T (Trial 2 production time - Trial 1 production time) and absolute values of $\Delta$T ($|$Trial 2 production time - Trial 1 production time$|$). The average $\Delta$T is positive, suggesting that there was an overall increase in time production in trial 2 as compared to trial 1. It was found that 64\% of participants increase their production time in trial 2 ($\Delta$T $ >$ 0). \\

\begin{center}
\begin{tabular}{lrr}

{} &  $\Delta$T &  absolute $\Delta$T \\
\hline
count &  986.0 &      986.0 \\
mean  &    2.7 &        7.0 \\
std   &   10.1 &        7.7 \\
min   &  -68.0 &        0.0 \\
25\%   &   -1.9 &        2.1 \\
50\%   &    2.4 &        4.7 \\
75\%   &    7.4 &        9.5 \\
max   &   49.2 &       68.0 \\
\hline
\end{tabular}
\end{center}

\subsubsection{Change in time production for each type of video engagement level combination}

There are a total of 9 video engagement level combinations for trial 1 and trial 2 videos. We check whether the $\Delta$T (difference between trial 2 and trial 1 time) for a particular combination significantly differs from the others. The transitions have been abbreviated using the first letter of each engagement level. For example, a transition from a low engagement video in trial 1 to a high engagement level video in trial 2 is denoted as L|H.  The following transitions were found to be significantly different from each other. 

\begin{itemize}
    \item H | L (Mean=1.6, STD=11.4) significantly different than M | H (Mean= 4.7, STD=10.3): t-statistic=-2.06, pvalue=0.04
    \item H | H (Mean=4, STD=9.8) significantly different than M | L (Mean= 1.4, STD=7.6): t-statistic=2.19, pvalue=0.03
    \item L | H (Mean=1.4, STD=10.2) significantly different than M | H (Mean= 4.7, STD=10.3): t-statistic=-2.34, pvalue=0.02
    \item M | L (Mean=1.4, STD=7.6) significantly different than M | H (Mean= 4.7, STD=10.3): t-statistic=-2.62, pvalue=0.009
\end{itemize}

For a high engagement level in trial 2 video the difference in time in significantly lower for a low engagement prior (trial 1 video) (Mean=1.4, STD=10.2) as compared to a medium engagement prior (Mean= 4.7, STD=10.3). This suggests that a high engagement video leads to a greater shift in attention away from time when preceded by a medium engagement video as compared to a low engagement video.  \\
Alternatively, for a medium engagement level in trial 1, the difference in time production is significantly lower when trail 2 video is low engagement level (Mean=1.4, STD=7.6) as compared to a high engagement video in trial 2 (Mean= 4.7, STD=10.3). This suggests that after watching a medium engagement video, a high engagement video leads to more distraction away from time as compared to a low engagement video. 

\noindent Table below shows the statistics of $\Delta$T for each engagement level combination. \\

\resizebox{\linewidth}{!}{\begin{tabular}{l c c c c c c c c c}
\hline
{} &  H$|$L &  L$|$M &  M$|$M & H$|$M &  H$|$H &  L$|$H &  M$|$L &  L$|$L &  M$|$H \\
\hline
count &                             108 &                               113&                                  113 &                                114 &                              108 &                             109 &                               109 &                            109 &                                103 \\
mean  &                               1.6 &                                 3.0 &                                    2.6 &                                  3.0 &                                4.0 &                               1.4 &                                 1.4 &                              2.5 &                                  4.7 \\
std   &                              11.4 &                                 7.9 &                                    9.6 &                                 13.2 &                                9.8 &                              10.2 &                                 7.6 &                              9.5 &                                 10.3 \\
min   &                             -68.0 &                               -18.9 &                                  -29.2 &                                -66.8 &                              -46.3 &                             -34.3 &                               -25.4 &                            -54.8 &                                -27.3 \\
25\%   &                              -2.7 &                                -1.5 &                                   -2.4 &                                 -1.3 &                               -0.5 &                              -2.9 &                                -2.9 &                             -0.6 &                                 -1.9 \\
50\%   &                               1.5 &                                 2.0 &                                    2.7 &                                  3.9 &                                2.8 &                               1.4 &                                 2.0 &                              2.7 &                                  3.1 \\
75\%   &                               6.3 &                                 6.8 &                                    6.9 &                                 10.9 &                                8.0 &                               5.4 &                                 6.3 &                              6.2 &                                 10.7 \\
max   &                              33.4 &                                34.1 &                                   49.2 &                                 31.9 &                               30.9 &                              40.4 &                                21.3 &                             33.9 &                                 38.9 \\
\hline
\end{tabular}}

\subsection{Video stimuli}
Links to videos can be made available upon request

\subsection{Experiment questionnaire}
\begin{figure}
    \centering
    \includegraphics[width=15cm, height=10cm]{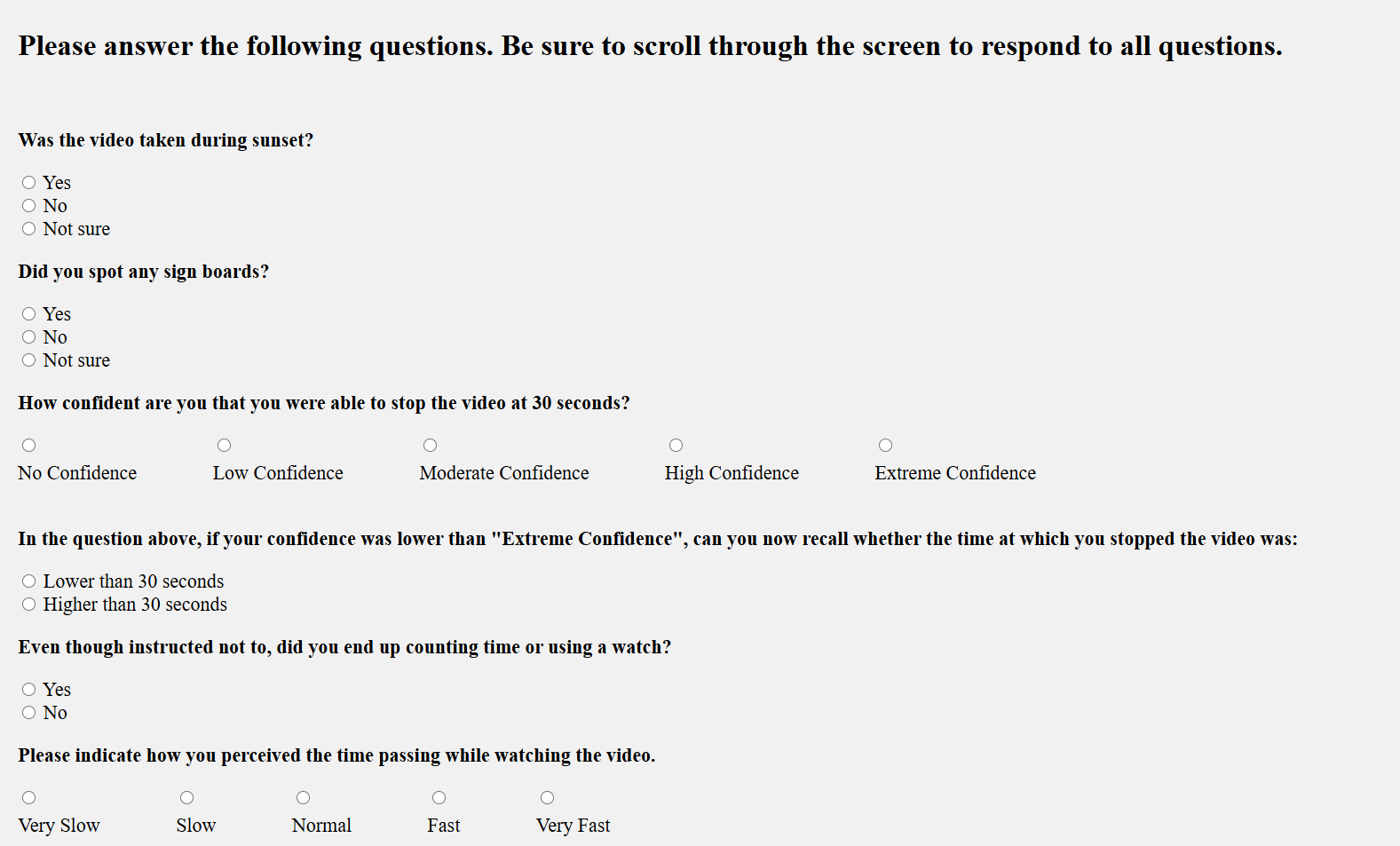}
\includegraphics[width=15cm, height=8cm]{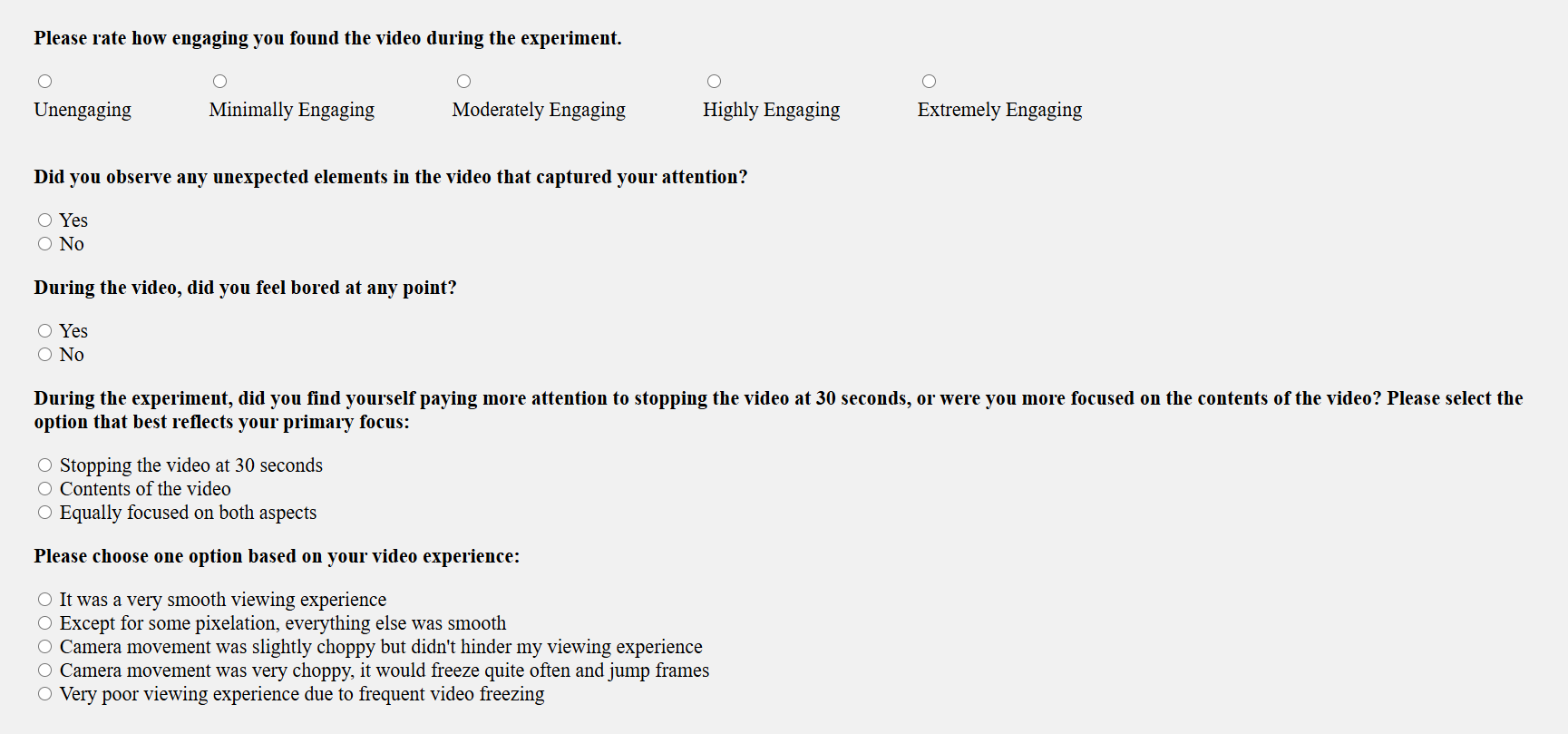}
    \caption{Questionnaire used after every trial in the main experiment}
    \label{fig:questionnaire}
\end{figure}

Refer Figure \ref{fig:questionnaire} for the questionnaire showed after the video in both the trials of the main experiment. 

\subsection{Feature selection}
A total of 13 features were extracted from different data points in the main experiment. Table below enlists all the 13 feature names. In addition to the features described in the main text, the following features were extracted: 

\begin{center}
\begin{tabular}{|c|}

\hline
Feature Names \\
\hline
 \textit{T1RelError} \\

 \textit{T1LowerThan30} \\
  ConfidenceIntiming \\
 
 PerceivedPassageOfTime \\
  PerceivedEngagmentLevel \\
  ReportedUnexpectednessInVideo \\
  ReportedBoredom \\
  \textit{HighVisualSensitivity} \\
  LowVisualSensitivity \\
  ReportedFocus \\
 V1EngagementLevel \\
  \textit{V2EngagementLevel} \\ 
  \textit{ChangeInEngagementLevel} \\

\hline
\end{tabular}
\end{center}

\begin{itemize}
    \item ConfidenceIntiming - Participants were asked how confident they were about stopping the video at 30 seconds (refer Figure \ref{fig:questionnaire}). The feature, ConfidenceIntiming, captures this information. 
    \item Participant responses to perception questions - In the questionnaire following each video   (refer Figure \ref{fig:questionnaire}), participants were asked about their perceived passage of time (PerceivedPassageOfTime), perceived engagement level of the video (PerceivedEngagmentLevel), whether they observed any unexpected elements in the video (ReportedUnexpectednessInVideo), whether they were bored (ReportedBoredom), and whether they focused on stopping the video or on video contents or both (ReportedFocus). Based on the responses, two additional features were derived. Namely, \textit{HighVisualSensitivity} (discussed in main text) and LowVisualSensitivity (which corresponds to participants reporting low engagement for a high engagement video). 
    \item  V1EngagementLevel, corresponds to the objective engagement level of video 1. 
\end{itemize}

\noindent Since there were only 986 samples to train, validate and test the model, we used feature elimination techniques to remove features that did not carry any useful information for prediction. We used permutation feature importance as the feature elimination technique. Permutation feature importance is a technique to measure the contribution of a single feature towards a fitted models performance. It involves shuffling the values of a specific feature and measuring how much the model's performance deteriorates as compared to when the feature values were not shuffled. 

In addition to good model performance the aim of this study was to explain the predictions using time perception theory. Thus, having as many features as possible  that can be linked to different components of the attentional-gate model was preferred. Furthermore, among the different features, the controllable features (features values that can be controlled by the experimenter, e.g. engagement level of trial 2 video) were given some preference since they represented the experiment stimulus and understanding how these features could contribute to the prediction model's output would be interesting. Hence, as long as the performance was not affected much, throughout the feature selection process we try to keep as many features as possible (especially the controllable ones). 

Table below shows the permutation feature importance of all 13 features for each of the trained models. The process for calculating permutation feature importance is as follows: For a given feature combination (all 13 features in this case), the best model parameters are obtained (Refer Model parameter tuning Section below for a detailed explanation on model parameter tuning process). The model is re-trained using the best parameters and permutation feature importance is calculated for the trained model using sklearn's "permutation\_importance" function. (The abbreviations used for model names in the tables of this section are as follow: Linear SVC - Linear support vector classifier, SVC - Support vector classifier, MLP - Multi-layer perceptron, NB - Naive Bayes)\\

\begin{center}
\resizebox{12cm}{3cm}{\begin{tabular}{c m{5em} m{3em} m{3em} m{3em} m{3em} m{3em}}
\hline
 & Logistic Regression & Random Forest & Linear SVC & SVC & MLP & NB \\
 \hline
\textbf{\textit{T1RelError}} & 0.07 & 0.08 & 0.08 & 0.05 & 0.08 & 0.03 \\
\textit{T1LowerThan30} & 0.01 & 0.01 & 0.01 & 0.01 & 0.02 & 0.02 \\
ConfidenceIntiming & 0.01 & 0.01 & 0.01 & 0.00 & 0.01 & 0.01 \\
\textit{HighVisualSensitivity} & 0.01 & 0.00 & 0.01 & 0.00 & 0.00 & 0.01 \\
PerceivedEngagmentLevel & -0.00 & 0.00 & -0.00 & -0.00 & 0.01 & -0.00 \\
ReportedUnexpectednessInVideo & 0.00 & -0.00 & 0.00 & -0.01 & 0.00 & 0.00 \\
\textit{ChangeInEngagementLevel} & -0.00 & -0.00 & -0.00 & 0.00 & 0.00 & -0.00 \\
\textit{V2EngagementLevel} & 0.00 & -0.00 & 0.00 & -0.01 & 0.01 & -0.00 \\
LowVisualSensitivity & 0.00 & -0.01 & 0.00 & -0.00 & -0.00 & 0.00 \\
FocusContent & -0.00 & -0.01 & -0.00 & -0.00 & 0.00 & -0.00 \\
FocusEqual & -0.00 & 0.00 & -0.00 & -0.00 & 0.00 & 0.00 \\
PerceivedPassageOfTime & -0.01 & 0.00 & -0.00 & -0.01 & 0.01 & 0.00 \\
ReportedBoredom & -0.01 & -0.00 & -0.00 & -0.01 & -0.00 & -0.00 \\
V1EngagementLevel & -0.01 & -0.00 & -0.01 & -0.02 & 0.00 & -0.00 \\
FocusStopping & -0.01 & -0.00 & -0.00 & -0.00 & 0.00 & -0.00 \\

\hline
\end{tabular}}
\end{center}

\noindent The feature permutation results reveal that \textit{T1RelError} is the most important feature and the performance (of all models) is not affected much by other features. It suggests that this feature is enough for having a good prediction score. 

Since the aim is to get as many features as possible, we remove the feature, \textit{T1RelError} and check the model performance. As expected the model performance drops, but it still performs better than a random model  (Refer the Table below in the section Model parameter tuning for results of different combinations of the features).  Feature permutation is again calculated for the new feature combination (all features excluding T1RelError) using the same process described above.  The results (See Table below) revealed that the feature, \textit{T1LowerThan30} is the most important feature in absence of the prior timing performance feature, \textit{T1RelError}. The second most important feature for most models, that had some contribution ($>$ 0.01 ) to the model predictions (in the absence of the prior timing performance feature) was \textit{HighVisualSensitivity}. Hence we consider these two features in the final set of selected features. 

\begin{center}
\resizebox{12cm}{3cm}{\begin{tabular}{c m{5em} m{3em} m{3em} m{3em} m{3em} m{3em}}
\hline
 & Logistic Regression & Random Forest & Linear SVC & SVC & MLP & NB \\
 \hline
\textbf{T1LowerThan30} & 0.038& 0.030& 0.064& 0.063& 0.036& 0.024
\\
ConfidenceIntiming & 0.005& 0.004& 0.000& 0.002& 0.007& 0.002
\\
 PerceivedPassageOfTime & -0.003& -0.004& 0.000& 0.003& -0.001&0.003
\\
PerceivedEngagmentLevel & 0.010& -0.007& 0.000& 0.001& 0.005& -0.002
\\
ReportedUnexpectednessInVideo & -0.003& -0.002& 0.000& -0.003& 0.003& -0.001
\\
 ReportedBoredom & 0.001& 0.000& 0.000& 0.003& -0.001&-0.007
\\
ChangeInEngagementLevel & 0.008& -0.002& 0.000& 0.001& 0.000& 0.003
\\
 V1EngagementLevel & 0.005& 0.000& 0.000& 0.002& 0.000&0.002
\\
V2EngagementLevel & 0.010& -0.003& 0.000& 0.002& 0.007& -0.002
\\
 \textbf{HighVisualSensitivity} & 0.013& 0.010& 0.001& 0.003& 0.011&0.016
\\
LowVisualSensitivity & -0.004& 0.002& 0.001& 0.000& 0.008& 0.011
\\
FocusContent & 0.005& 0.004& 0.000& 0.002& 0.001& 0.000
\\
FocusEqual & -0.002& 0.000& 0.000& -0.001& -0.005& 0.001
\\
FocusStopping & -0.002& 0.000& 0.000& -0.001& -0.005& 0.001
\\
\hline
\end{tabular}}
\end{center}

Additionally the features \textit{V2EngagementLevel} and ChangeInEngagementLevel can be classified as controllable features. This is because, the values of these features can be controlled by the experimenter.  Since their addition does not affect the final model performance, we add these features to the list of selected features. It is important to note that the two features are dependent.  For example, when \textit{ChangeInEngagementLevel} equals 0, this indicates that the engagement level of the trial 2 video must be lower than that of the trial 1 video. Consequently, \textit{V2EngagementLevel} in this case cannot have a value of 2 (which corresponds to a high engagement level) because the engagement level of the trial 1 video would need to be higher than that of trial 2, which is not possible. Therefore, the two video characteristic features are restricted from taking opposite extreme values—one cannot be 0 while the other is 2 and vice versa. Note that V1EngagementLevel, is also controlled by the experimenter. However, it corresponds to the first trial's video engagement level. Since the model predicts only after the first trial, this feature is considered to be in the past and cannot be changed or controlled. That is why this feature is not considered a controllable feature.  

\subsection{Model parameter tuning }

Table below shows model performance for different feature combinations. The test score has been obtained using the following procedure. First, data is undersampled to have the same number of increase and decrease cases. Then using 5-fold cross validation with f1-score as a metric, the best model parameters are found using GridSearch function from sklearn in python. The overall results from the best fitted model are considered. This is repeated 3 time for different splits of the data (due to under-sampling). The results in the table below represent the overall mean for all the cross-validation test sets and the 3 iterations. We can see that, in most cases the best performance is using the prior timing performance feature, \textit{T1RelError} alone. However as discussed above, since other features come into play when \textit{T1RelError} is ignored and since the final performance is not affected much by including the other features we finalized on the five features discussed in the main text. In the local feature analysis section we see how the different features interplay when prior timing performance feature does not contribute much to the final prediction. 

\resizebox{\linewidth}{!}{\begin{tabular}{|c|m{5em}|m{5em}|m{5em}|m{5em}|}
\hline
Model name  &  All features &  \textit{T1RelError} only &  \textit{T1RelError} excluded &  Selected features \\
\hline
Support vector classifier &  0.59 &  0.60 &  0.56 &  0.58 \\
Random Forest &  0.60 &  0.58 &  0.55 &  0.59 \\
Naive Bayes &  0.56 &  0.53 &  0.52 &  0.56 \\
Multi-layer perceptron &  0.59 &  0.61 &  0.56 &  0.60 \\
Logistic Regression &  0.59 &  0.60 &  0.55 &  0.59 \\
Linear Support vector classifier &  0.59 &  0.60 &  0.56 &  0.60 \\

\hline
\end{tabular}}\\

\resizebox{\linewidth}{!}{\begin{tabular}{|m{5em}|m{5em}|m{5em}|m{5em}|m{5em}|}
\hline
Model name  &  All features &  \textit{T1RelError} only &  \textit{T1RelError} excluded &  Selected features \\
\hline
Support vector classifier & 'C': 14.18, 'degree': 2, 'kernel': 'linear' & 'C': 28.41, 'degree': 2, 'kernel': 'rbf' & 'C': 0.29, 'degree': 2, 'kernel': 'sigmoid' &  'C': 28.41, 'degree': 2, 'kernel': 'rbf' \\
\hline
Random Forest & 'bootstrap': True, 'max depth': 4, 'min samples split': 14 & 'bootstrap': True, 'max depth': 8, 'min samples split': 6 & 'bootstrap': False, 'max depth': 4, 'min samples split': 14 &  'bootstrap': True, 'max depth': 4, 'min samples split': 20  \\
\hline
Naive Bayes &  &  &  &   \\
\hline
Multi-layer perceptron & 'activation': 'tanh', 'alpha': 0.1, 'hidden layer sizes': (80, 40), 'solver': 'sgd' & 'activation': 'relu', 'alpha': 0.1, 'hidden layer sizes': (80, 40), 'solver': 'adam' & 'activation': 'tanh', 'alpha': 0.05, 'hidden layer sizes': (80,), 'solver': 'sgd' &  'activation': 'relu', 'alpha': 0.1, 'hidden layer sizes': (80, 40), 'solver': 'sgd' \\
\hline
Logistic Regression & 'C': 0.14, 'solver': 'liblinear' & 'C': 12.06, 'solver': 'lbfgs' & 'C': 7.57, 'solver': 'lbfgs' &  'C': 12.06, 'solver': 'lbfgs' \\
\hline
Linear support vector classifier & 'C': 0.53, 'loss': 'squared hinge', 'penalty': 'l2' & 'C': 0.80, 'loss': 'squared hinge', 'penalty': 'l2' & 'C': 0.69, 'loss': 'hinge', 'penalty': 'l2' &  'C': 3.66, 'loss': 'squared hinge', 'penalty': 'l2' \\

\hline
\end{tabular}}\\
\\
\noindent The final model parameters for the different feature combinations are shown above.

\subsection{Machine learning model selection}

The table below shows the performance of all models for direction of change prediction using LOOCV (as in the main text).  The logistic regression model, linear support vector classifier and multi-layer perceptron show the best accuracy. 

\begin{center}
\begin{tabular}{m{10em} m{4em} m{3em} m{3em} }
\hline
                           Model &  precision &  recall &  accuracy\\
  \hline
        Logistic Regression &  0.59 &  0.59 &          0.59 \\
        \hline
              Random Forest &  0.56 &  0.56 &          0.56 \\
              \hline
                 Linear Support vector classifier &  0.59 &  0.58 &          0.59 \\
                 \hline
                        Support vector classifier &  0.54 &  0.45 &          0.53 \\
                        \hline
                        Multi-layer perceptron &  0.58 &  0.57 &          0.58 \\
                        \hline
                         Naive Bayes &  0.56 &  0.59 &          0.56 \\
\hline
\end{tabular} 
\end{center}

The model coefficients (for logistic regression models and linear support vector machine) and feature importance (for random forest model) when the models are trained using LOOCV are given in the table below. The coefficients can be considered same as feature importance since the features were scaled using StandardScalar function from pyhton's sklearn. 

The logistic regression model was selected for this study. The reasons for selecting the logistic regression model was it's simplicity compared to random forest and Multi-layer perceptron, implicit probability calibration  unlike Naive Bayes and a balanced distribution of feature importance across all selected features. Compared to other models, the logistic regression model has fewer features with almost zero importance (especially for the controllable features, \textit{ChangeInEngagementLevel} and \textit{V2EngagementLevel}). Additionally, the implicit probability calibration of the model may imply that the output probabilities could convey additional insights beyond merely predicting the direction of change. 

\begin{center}
\begin{tabular}{c m{5em} m{4em} m{2em}}
\hline
           feature name &  Logistic Regression &  Random Forest &  Linear SVC \\
\hline
          \textit{T1LowerThan30} &                -0.16 &           0.10 &       -0.07 \\
          \hline
\textit{ChangeInEngagementLevel} &                 0.08 &           0.05 &        0.04 \\
\hline
      \textit{V2EngagementLevel} &                -0.16 &           0.08 &       -0.08 \\
      \hline
  \textit{HighVisualSensitivity} &                -0.22 &           0.07 &       -0.10 \\
  \hline
             \textit{T1RelError} &                 0.50 &           0.70 &        0.23 \\
             \hline
              intercept &                 0.01 &           0.00 &        0.00 \\
              \hline

\end{tabular}
\end{center}

\subsection{Local feature analysis results at sub-group level}

The table below shows aggregated SHAP values of the main feature combinations discussed in the "Cognitive interpretation of individual predictions" section in the main text. The table also includes the model accuracy , average probability and counts for the corresponding subset. 
The threshold for extreme values of trial 1 production time was decided roughly based on percentiles. The 25th and 75th percentile of trial 1 production time are 27 seconds and 38 seconds respectively. The low values of this feature are defined to be below 25 seconds (less than 25 percentile) and high values to be above 45 seconds (more than 75 percentile).
 We found that for extreme cases of trial 1 production time, the model solely basis it's predictions on the prior timing features (as seen from the SHAP values of different features in the table below). The accuracy and average probability of the model confirm that it effectively predicts the probability of a decrease in time perception with high confidence when prior timing features take extreme values. Thus, the accuracy of the model for each subset confirms that some patterns detected in the data by the model are effective in predicting the direction of change in time production. These are, 1. Time production will increase if previous trial's production time is very low and vice versa. 2. Participants marked as sensitive tend to increase time production in the next trial. \\

\begin{center}
\resizebox{12cm}{2.5cm}{\begin{tabular}{ c m{5em} m{5em} m{5em} m{10em}}
\hline
Feature name & Production $<$ 25 seconds  & Production $>$ 45 seconds & Production Bet(25,45) & \textit{HighVisualSensitivity}=1 (all Production values) \\
\hline
\textit{T1LowerThan30} & -0.05 & 0.04 & -0.01 & -0.07 \\
\textit{ChangeInEngagementLevel} & 0.00 & -0.01 & -0.02 & 0.12 \\
\hline
\textit{V2EngagementLevel} & -0.02 & -0.02 & 0.02 & 0.00 \\
\hline
\textit{HighVisualSensitivity} & 0.05 & 0.03 & 0.03 & -0.94 \\
\hline
\textit{T1RelError} & -0.74 & 1.16 & -0.05 & 0.08 \\
\hline
\textbf{Accuracy} & 0.67 & 0.72 & 0.57 & 0.73 \\ 
\hline
\textbf{Average probability}  & 0.33 & 0.75 & 0.49 & 0.31 \\
\hline
\textbf{Count} & 132 & 102 & 472 & 41 \\
\hline
\end{tabular}}
\end{center}

\end{document}